\documentclass[twocolumn]{aastex631}
\usepackage{bm}
\usepackage{amsmath}
\usepackage{booktabs}
\usepackage{tabularx}
\usepackage{array}

\begin{document}

\title{Exploring cosmological constraints of the weak gravitational lensing and galaxy clustering joint analysis in the CSST photometric survey}

\correspondingauthor{Yan Gong}
\email{Email: gongyan@bao.ac.cn}

\author{Qi Xiong}
\affiliation{National Astronomical Observatories, Chinese Academy of Sciences, 20A Datun Road, Beijing 100101, China}
\affiliation{School of Astronomy and Space Sciences, University of Chinese Academy of Sciences(UCAS),\\Yuquan Road NO.19A Beijing 100049, China}

\author[0000-0003-0709-0101]{Yan Gong*}
\affiliation{National Astronomical Observatories, Chinese Academy of Sciences, 20A Datun Road, Beijing 100101, China}
\affiliation{School of Astronomy and Space Sciences, University of Chinese Academy of Sciences(UCAS),\\Yuquan Road NO.19A Beijing 100049, China}
\affiliation{Science Center for China Space Station Telescope, National Astronomical Observatories, Chinese Academy of Science, \\20A Datun Road, Beijing 100101, China}

\author{Xingchen Zhou}
\affiliation{National Astronomical Observatories, Chinese Academy of Sciences, 20A Datun Road, Beijing 100101, China}
\affiliation{Science Center for China Space Station Telescope, National Astronomical Observatories, Chinese Academy of Science, \\20A Datun Road, Beijing 100101, China}

\author{Hengjie Lin}
\affiliation{National Astronomical Observatories, Chinese Academy of Sciences, 20A Datun Road, Beijing 100101, China}

\author{Furen Deng}
\affiliation{National Astronomical Observatories, Chinese Academy of Sciences, 20A Datun Road, Beijing 100101, China}
\affiliation{School of Astronomy and Space Sciences, University of Chinese Academy of Sciences(UCAS),\\Yuquan Road NO.19A Beijing 100049, China}
\affiliation{Institute of Astronomy, University of Cambridge, Madingley Road, Cambridge, CB3 0HA, UK}

\author{Ziwei Li}
\affiliation{South-Western Institute for Astronomy Research, Yunnan University, Kunming 650500, China}

\author[0000-0002-0966-8598]{Ayodeji Ibitoye}
\affiliation{Department of Physics, Guangdong Technion - Israel Institute of Technology, Shantou, Guangdong 515063, China}
\affiliation{Centre for Space Research, North-West University, Potchefstroom 2520, South Africa}
\affiliation{National Astronomical Observatories, Chinese Academy of Sciences, 20A Datun Road, Beijing 100101, China}
\affiliation{Department of Physics and Electronics, Adekunle Ajasin University, P. M. B. 001, Akungba-Akoko, Ondo State, Nigeria}

\author{Xuelei Chen}
\affiliation{National Astronomical Observatories, Chinese Academy of Sciences, 20A Datun Road, Beijing 100101, China}
\affiliation{School of Astronomy and Space Sciences, University of Chinese Academy of Sciences(UCAS),\\Yuquan Road NO.19A Beijing 100049, China}
\affiliation{Department of Physics, College of Sciences, Northeastern University, Shenyang 110819, China}
\affiliation{Centre for High Energy Physics, Peking University, Beijing 100871, China}

\author{Zuhui Fan}
\affiliation{South-Western Institute for Astronomy Research, Yunnan University, Kunming 650500, China}

\author{Qi Guo}
\affiliation{National Astronomical Observatories, Chinese Academy of Sciences, 20A Datun Road, Beijing 100101, China}
\affiliation{School of Astronomy and Space Sciences, University of Chinese Academy of Sciences(UCAS),\\Yuquan Road NO.19A Beijing 100049, China}

\author{Ming Li}
\affiliation{National Astronomical Observatories, Chinese Academy of Sciences, 20A Datun Road, Beijing 100101, China}

\author{Yun Liu}
\affiliation{National Astronomical Observatories, Chinese Academy of Sciences, 20A Datun Road, Beijing 100101, China}
\affiliation{School of Astronomy and Space Sciences, University of Chinese Academy of Sciences(UCAS),\\Yuquan Road NO.19A Beijing 100049, China}

\author{Wenxiang Pei}
\affiliation{National Astronomical Observatories, Chinese Academy of Sciences, 20A Datun Road, Beijing 100101, China}
\affiliation{School of Astronomy and Space Sciences, University of Chinese Academy of Sciences(UCAS),\\Yuquan Road NO.19A Beijing 100049, China}



\begin{abstract}

We explore the joint weak lensing and galaxy clustering analysis from the photometric survey operated by the China Space Station Telescope (CSST), and study the strength of the cosmological constraints. We employ a high-resolution JiuTian-1G simulation to construct a partial-sky light cone to $z=3$ covering 100 deg$^2$, and obtain the CSST galaxy mock samples based on an improved semi-analytical model. We perform a multi-lens-plane algorithm to generate corresponding synthetic weak lensing maps and catalogs. Then we generate the mock data based on these catalogs considering the instrumental and observational effects of the CSST, and use the Markov Chain Monte Carlo (MCMC) method to perform the constraints. The covariance matrix includes non-Gaussian contributions and super-sample covariance terms, and the systematics from intrinsic alignments, galaxy bias, photometric redshift uncertainties, shear calibration, and non-linear effects are considered in the analysis.  We find that the constraint result is comparable to that from Stage III surveys, and it can be significantly improved further in the full CSST survey with 17500 deg$^2$. This indicates the CSST photometric survey is powerful for exploring the Universe.

\end{abstract}

\keywords{cosmological parameters --- large-scale structure of universe --- dark matter}


\section{Introduction} \label{sec:intro}

Our understanding of the Universe has been greatly improved in the past decades due to more and more accurate cosmological observations, such as the cosmic microwave background (CMB; e.g \cite{planck18}),  Type Ia supernovae (e.g \cite{typeIa1,typeIa2}), and baryon acoustic oscillations (BAO; e.g \cite{BAO1}). Analysis from these observations has statistically converged, and a standard cosmological model characterized by a small number of parameters, i.e. the $\Lambda$CDM model, has emerged \citep{LCDM1,LCDM2,planck18}. This model describes a spatially flat Universe composed of roughly $30\%$ matter (visible matter and cold dark matter (CDM)) and $\sim 70\%$ dark energy. This dark energy is responsible for the accelerating expansion of the Universe and is consistent with the behavior of the cosmological constant ($\Lambda$), whose physical nature is still poorly understood.

Cosmic large-scale structure (LSS) is one of the effective probes for studying the nature of dark energy and dark matter, which usually can be measured by the spatial distribution of galaxies, i.e. galaxy clustering. However, galaxy is a biased tracer of the underlying matter distribution. On the other hand, weak gravitational lensing, also known as cosmic shear, is a powerful technique for exploring the matter distribution in the Universe without assuming a specific correlation between dark matter and baryons \citep{1992ApJ...388..272K,2000A&A...358...30V}. 

Recently, combinations of galaxy clustering and weak lensing (so-called 3$\times$2pt probes) have been widely used for extracting cosmological information from photometric galaxy surveys \citep{3x2pt1,3x2pt2,3x2pt3,3x2pt4,3x2pt5,3x2pt6}, such as the Dark Energy Survey\footnote{\url{https://www.darkenergysurvey.org/}} (DES), Kilo Degree Survey\footnote{\url{https://kids.strw.leidenuniv.nl/}} (KiDS), and Hyper Suprime Cam\footnote{\url{https://www.naoj.org/Projects/HSC/index.html}} (HSC). In particular, we can extract accurate information from a joint analysis of galaxy clustering, weak lensing and galaxy-galaxy lensing from the configuration/real space angular correlation functions \citep{DESreal1,DESreal2}, or using their Fourier/Harmonic counterpart, i.e. the two-point angular power spectrum \citep{KiDsfour1,KiDsfour2,KiDsfour3}. These ongoing Stage~III surveys have played a crucial role in testing the $\Lambda$CDM model and have provided effective constraints on several important cosmological parameters related to dark energy and dark matter \citep[see e.g.][]{2022PhRvD.106d3520P,2021A&A...646A.140H,2023OJAp....6E..36D,2019PASJ...71...43H,2020PASJ...72...16H}.

In the next decade, the upcoming Stage~IV surveys, such as Vera C. Rubin Observatory\footnote{\url{https://rubinobservatory.org/}} \citep[or LSST,][]{LSST}, Nancy Grace Roman Space Telescope\footnote{\url{https://roman.gsfc.nasa.gov/}} \citep[or WFIRST,][]{WFIRST}, {\it Euclid}\footnote{\url{https://www.euclid-ec.org/}}\citep{Euclid}, and China Space Station Telescope \citep[CSST,][]{zhan1,zhan2,zhan3,gong}, will explore enormous volumes of the Universe and enable us to constrain the cosmological parameter to an unprecedented precision. 
The CSST is a 2-meter space telescope in the same orbit of the China Manned Space Station, and it can simultaneously perform photometric imaging and slitless spectroscopic surveys covering $\sim$ 17,500 deg$^2$ survey area with high spatial resolution and wide wavelength coverage. It has seven photometric imaging bands (i.e $NUV$,$u,g,r,i,z$ and $y$) and three spectroscopic bands (i.e $GU$, $GV$ and $GI$) ranging from 250 to 1000~nm. The CSST photometric survey can reach a magnitude limit $i\sim 26.0$ for $5\sigma$ point source detection. 

In this paper, we explore the cosmological constraint power of the CSST 3$\times$2pt survey based on a 100 deg$^2$ area using simulations. We construct the mock galaxy catalog using an updated version of the L-Galaxies semi-analytical model \citep{henriques2015galaxy,2024MNRAS.529.4958P}, and consider the CSST instrumental design and strategy of the CSST photometric survey. To reduce the repeated structures, we use special viewing angles to build the past light cones and perform the multi-lens-plane algorithm to generate a weak lensing catalog. We derive the galaxy redshift distribution from the mock catalog and divide it into several photometric-redshift (photo-$z$) tomographic bins. Then we compute the mock CSST two-point angular power spectra, including weak lensing, galaxy clustering, and galaxy-galaxy lensing. Finally we perform the constraints on cosmological and systematic parameters using the Markov Chain Monte Carlo (MCMC) method.

The paper is organized as follows: in Section \ref{sec:data} we briefly describe the gravity-only simulations and the light-cone construction, and introduce the generation of galaxy catalog. In Section \ref{map mk}, we show the details of the weak lensing catalog construction and the map-making procedure. In Section \ref{pk}, we estimate the galaxy clustering, cosmic shear, and galaxy-galaxy lensing power spectrum from the mock maps, and describe the theoretical modeling of the angular power spectrum including relevant systematics. The model fitting method and the discussion of constraint result of the cosmological and systematic parameters are given in Section \ref{fitting}, and our conclusions are summarized in Section \ref{summary}. 

\section{Mock Catalog} \label{sec:data}


\subsection{Simulation} \label{sec:sim}
The box size and the number of particles are two crucial parameters in cosmological simulations. For the upcoming Stage~IV surveys, simulations with volume $V \gtrsim $ $1h^{-3}$Gpc$^3$ and mass resolution of $m_{\rm p}$ $\sim$ $10^9$ $h^{-1}M_{\odot}$ are required to ensure the percent-level convergence of different measurements and statistical errors \citep{schneider2016matter,2019MNRAS.489.1684K}. Therefore we employ the JiuTian-1G (hereafter JT1G) simulation to serve as the basis to derive the weak lensing convergence map and mock galaxy catalog. 

JT1G is one of high-resolution simulations in the state-of-the-art N-body simulation suite JiuTian Simulation. Utilizing the L-Gadget3 code \citep{2005MNRAS.364.1105S}, the JT1G simulation tracks $6144^3$ dark matter particles within a cubic simulation box. The box size is $1h^{-1}$Gpc and the particle mass is about 3.72$\times$10$^8$ $h^{-1}$ $M_{\odot}$. The cosmological parameters are from $\it Planck$2018 \citep{planck18}, with $\Omega_{\text{m}} = 0.3111$, $\Omega_{\text{b}} = 0.0490$, $\Omega_\Lambda = 0.6899$,  $\sigma_8 = 0.8102$, $n_{\text{s}} = 0.9665$, and $h = 0.6766$. The JT1G simulation begins at initial redshift $z_{\rm ini} = 127$, and 128 snapshots are outputted between $z_{\rm ini }$ and $z=0$ with an average time gap of $\sim 100$ Myr, which is suitable for weak lensing studies. The dark matter halos and subhalos are identified using the friend-of-friends(FOF) and SUBFIND algorithm \citep{2001NewA....6...79S,2005MNRAS.364.1105S}.

\subsection{Light-cone Construction} \label{lightcone}
Even with a comoving size $L=1\,h^{-1}$Gpc, the simulation box is not big enough to trace back the light cone to the desired redshift within one box. One approach is to exploit the periodic boundary conditions of the simulation by arranging replicas of the simulation box along its preferred axis. However, this operation can introduce artificial structures known as the box replication effect, which manifests as a kaleidoscope-like pattern in the lensing map when the simulated box is small or the redshift is relatively high \citep{2005MNRAS.360..159B,2007MNRAS.376....2K,cz}. Results in \cite{cz} have shown that tilling simulation box at different redshift along special directions (e.g. the main axis directions, the diagonal directions of the x-y/y-z/x-z planes, and the diagonal orientations of cubic boxes) to construct partial-sky light cone would lead to many times of box copies, and introduce significant systematic bias in the weak lensing statistics.

In order to reduce repeated structures and take advantage of the periodicity of the simulation box, we choose to select a special viewing angle as the line-of-sight (LOS) direction to construct a partial-sky light cone to $z = 3$ \citep{2007MNRAS.376....2K,2009A&A...499...31H,2010ApJS..190..311C}, covering $100$ deg$^2$ sky area. The light cone is made up of a set of discrete slices based on the outputting snapshots at different redshifts in the redshift range of the simulation box. Specifically, the dark matter particles in the simulation box are assigned to each slice at a comoving distance $\chi$ using the snapshot with the nearest redshift $z_{\text{snap}} \sim z(\chi)$ to capture the evolution of redshift. Then we convert the comoving positions of particles in the simulation box into corresponding comving positions in the light cone. Using this method, artificial features disappear and unbiased construction can be achieved \citep{cz}, especially in the case when the opening angle of the light cone is small. 

\begin{figure}
\includegraphics[width=1\linewidth]{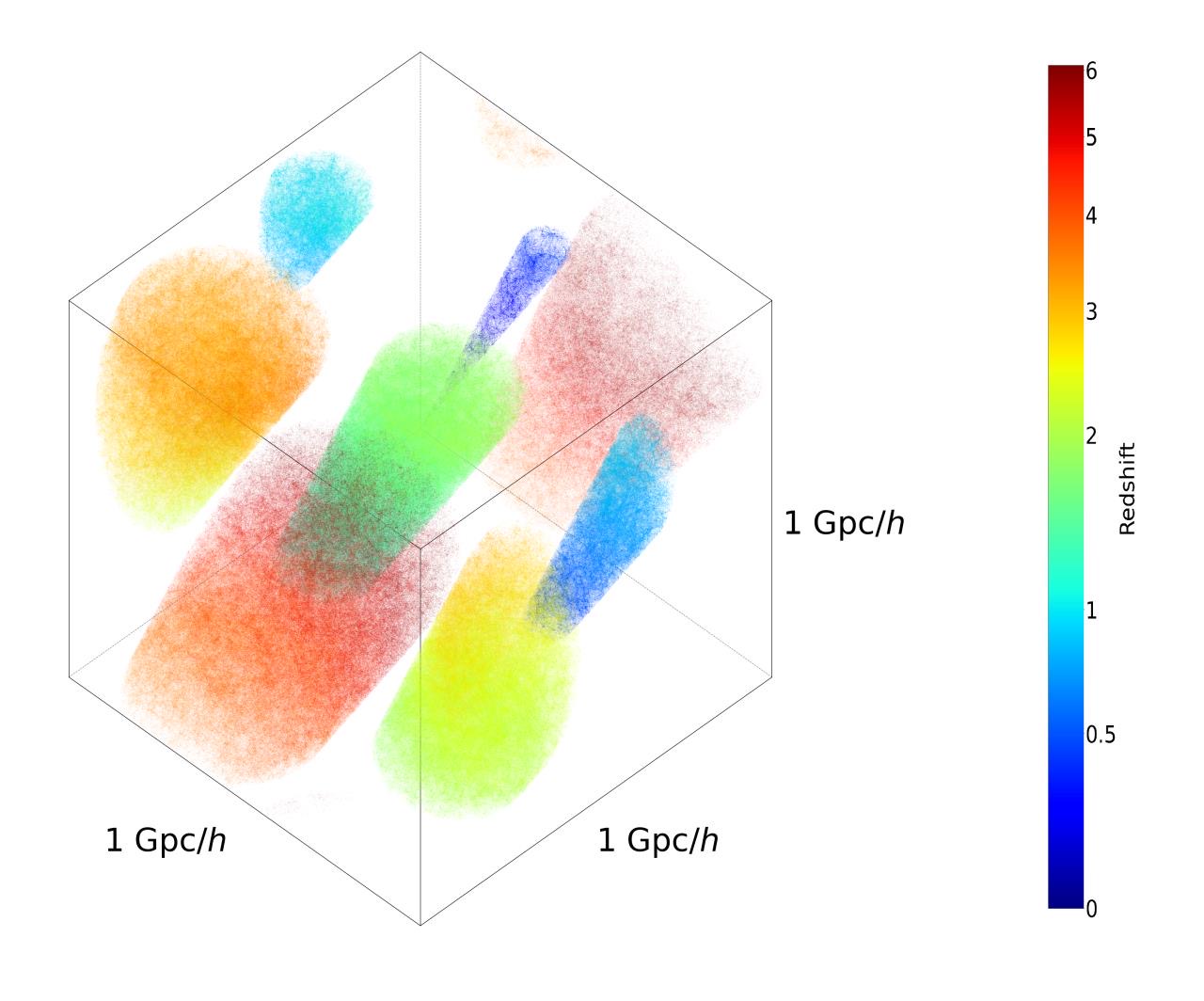}
\caption{Schematic illustrations of our partial-sky light-cone construction. Different colors of cones show the covered geometry in snapshots at different redshfits.}
\label{fig:lc}
\end{figure}
Figure \ref{fig:lc} simply illustrates the strategy of our partial-sky light-cone construction. Different colors of cones show the covered geometry in snapshots at different redshifts. Both dark matter particles and halos (or galaxies) light-cone construction follow the same strategy, so that different tracers capture the same cosmological information.

\subsection{Galaxy mock catalog} \label{gal}
The mock galaxy catalog is built by implementing an improved Semi-Analytic Model \citep{henriques2015galaxy,2024MNRAS.529.4958P}. In addition to the basic properties of galaxies, such as star-formation rate, stellar mass, and stellar metallicity, the mock catalog also provides observer-frame dust-corrected luminosities of the CSST photometry bands in the AB-magnitude system. These luminosities are determined by modeling the spectral energy distribution (SED) of a galaxy at redshift $z$ with the Stellar Population Synthesis (SPS) model \citep{BC03} and the Chabrier initial mass function \citep{Chabrier}. We integrate the resulting SED with dust-correction \citep{Devriendt,Charlot2000ApJ...539..718C,henriques2015galaxy} over the total throughput for the CSST filters, including filter intrinsic transmission, detector quantum efficiency, and mirror efficiency. Then the observer-frame luminosities are converted to the apparent magnitudes by
\begin{equation}
    m_{\rm AB} = -2.5\,\text{log} \left [\frac{(1+z)L_{R}}{4\pi r^2_{L}} \right],
\end{equation}
where $z$ and $r_{L}$ are the redshift and the luminosity distance of a galaxy, and $L_{R}$ is the observer-frame dust-corrected luminosity of a CSST band $R$. The factor ($1+z$) accounts for the compression of photon frequencies in the observer frame. We then use apparent magnitudes to select galaxies that can be detected by the CSST photometric survey. Specifically, we use the magnitude limit $i \leq 24.6$ ($5\sigma$ detection for extended sources)  to obtain the galaxy samples \citep{gong}. 

\begin{figure}
    \centering
    \includegraphics[width=1\linewidth]{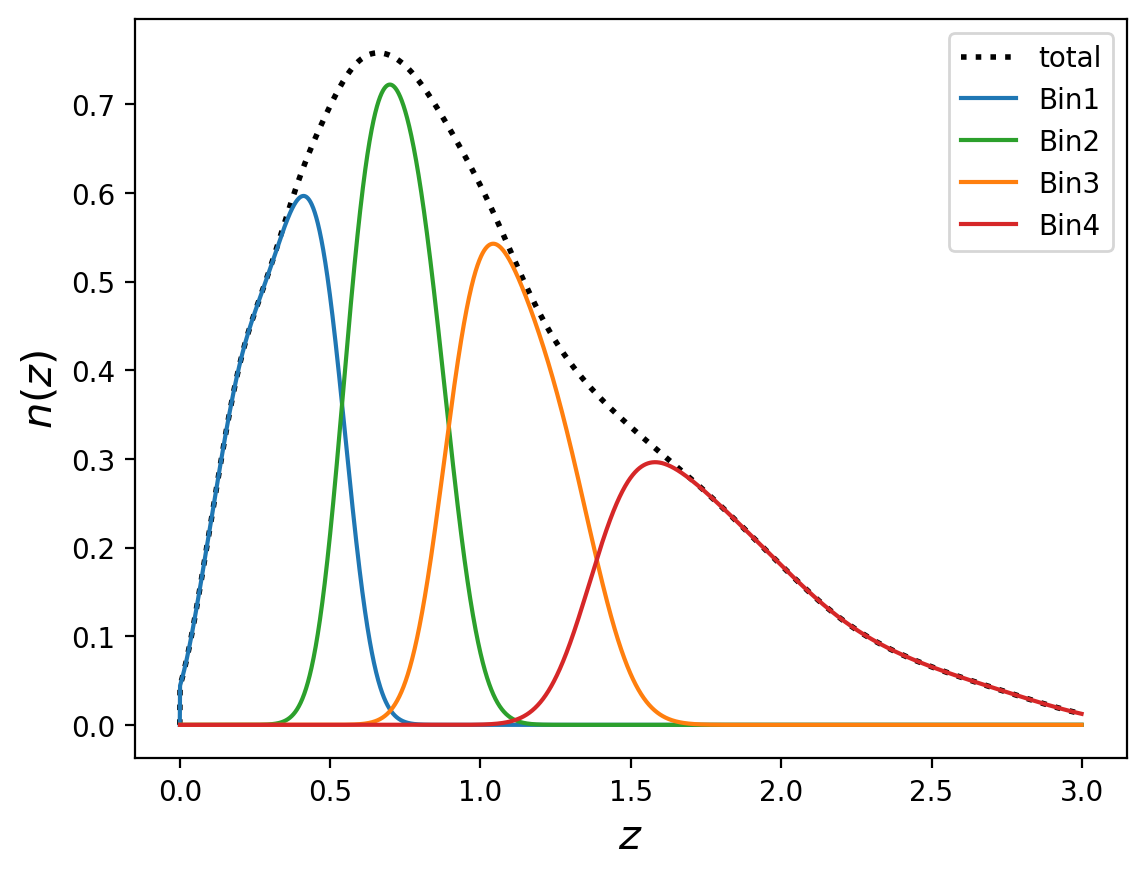}
    \caption{The mock galaxy redshift distributions in the CSST photometric survey. The black dotted curve shows the total redshift distribution $n(z)$, which is obtained by stacking samples from the redshift PDF of each individual galaxy. The solid blue, green, orange, and red curves denote the redshift distribution $n^{i}(z)$ of the four photo-$z$ bins.}
    \label{fig:red_dis}
\end{figure}

In order to obtain the photometric redshifts, we assume that the observed redshift fellows a Gaussian probability distribution function (PDF) for each individual galaxy with $z_{\text{obs}} \sim N(z_{\text{true}}, \sigma_{z})$, where $z_{\text{true}}$ is the true redshift and $\sigma_{z}$ is the redshift uncertainties characterized as $\sigma_{z} = \sigma_{z_0}(1+z)$.
Here, we assume $\sigma_{z_0}$ is a constant with $\sigma_{z_0} = 0.05$, which is the same as that in \cite{gong}.
The total photo-z redshift distribution derived from mock catalog is shown in Figure \ref{fig:red_dis} (black dotted curve), computed by stacking samples from the redshift PDF of each individual galaxy. We can find that the redshift distribution has a peak around $z = 0.6$, similar with the distribution derived from COSMOS catalog \citep{2007ApJS..172...99C,2009ApJ...690.1236I}, which can be used to represent the redshift and magnitude distributions observed by the CSST photometric survey \citep{cao,gong}.

\setlength{\tabcolsep}{14pt}
\renewcommand{\arraystretch}{1.5}
\begin{table}[h!]
        \centering
        \caption{The CSST samples used in this work. The photo-$z$ tomographic bins, number of galaxies in each bin ($N_{\text{gal}}$), the average angular number density of galaxies in arcmin$^{-2}$($\bar{n}_{g})$, and the shape noise per component ($\sigma_{\gamma}$) have been listed.}
        \label{tab:1}
        \begin{tabular}{cccc}
        \hline
        \hline
            redshift bin & $N_{\text{gal}}$ & $\bar{n}_{g}$ & $\sigma_{\gamma}$ \\
            \hline
         $0<z_{\text{ph}}<0.55$ & 2391071 & 6.64 & 0.2 \\
         $0.55<z_{\text{ph}}<0.89$ & 2384034 & 6.62 & 0.2 \\
         $0.89<z_{\text{ph}}<1.38$ & 2406640 & 6.69 & 0.2 \\
         $1.38<z_{\text{ph}}<3$ & 2365087 & 6.57 & 0.2 \\
        \hline
        \hline
    \end{tabular}

\end{table}

In order to extract more information from the weak lensing data, we divide the redshift range into different photo-z tomographic bins, and study the auto and cross power spectra of these bins. As shown in Figure \ref{fig:red_dis}, we split the galaxy samples into four tomographic redshift bins containing similar numbers of galaxies as shown in Table~\ref{tab:1}. We then estimate the redshift distribution in each tomographic bin by stacking all the redshift PDFs. Note that the fitting results of the cosmological parameters also depend on the number of photo-$z$ bins and more number of bins may improve the constraint results \citep{2002PhRvD..65f3001H}. (e.g. a factor of $\sim$ 1.5 for the six photo-z bins cases shown in \cite{gong}).
In the current stage, four tomographic bins are adequate for this work, and we use the same galaxy distribution for both lenses and sources, since it can help to constrain the photo-$z$ error \citep{2020JCAP...12..001S}. The catalog contains about 9,600,000 galaxies in 100 deg$^2$, and the average surface galaxy density $\sim$ 26.7 arcmin$^2$, which is consistent with the results for the CSST photometric survey derived from the COSMOS catalog \citep{gong}. 

\section{Map Making} \label{map mk}

\begin{figure*}
    \centering
    \includegraphics[width=1 \linewidth]{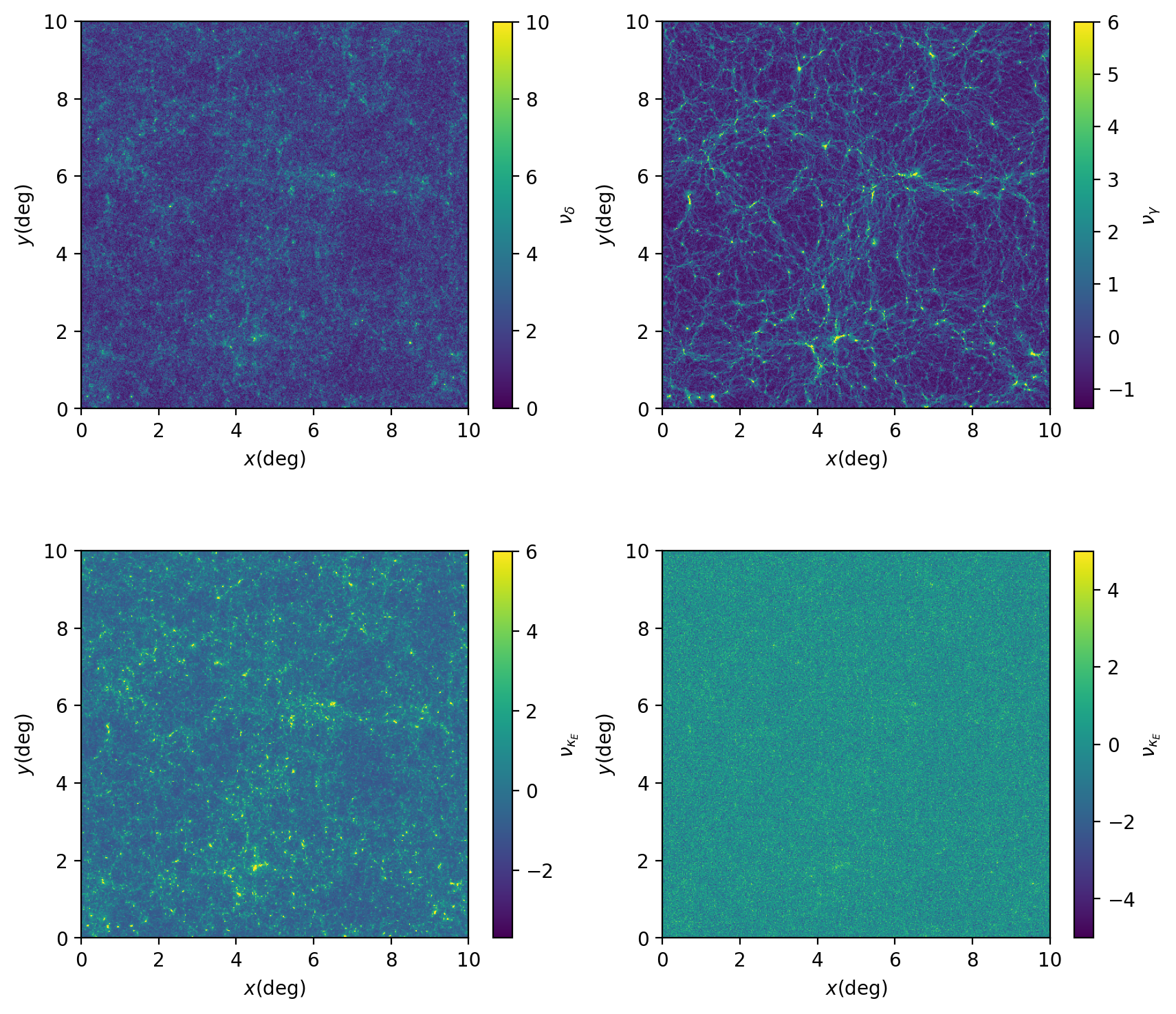}
    \caption{The simulated galaxy density and weak lensing maps in 100 deg$^2$ for the CSST photometric survey. For visualization purposes, we choose the first and second tomographic bins to show the galaxy density map and lensing maps, respectively. The upper left panel is the galaxy density map of the first tomographic bin, and the upper right panel shows the absolute value of the shear field of the second tomographic bin, obtained by combining the two components of the shear as, $\gamma = \sqrt{\gamma_1^2 + \gamma_2^2}$. The bottom left panel shows the KS-inversion method reconstructed convergence map, and the bottom right panel is the convergence map after adding the Gaussian shape noise. The color bars represent the fluctuation field in units of the field standard deviation, e.g $\nu_{\gamma} = (\gamma - \left < \gamma \right >)/\sqrt{(\gamma - \left < \gamma \right >) ^2}$.}
    \label{fig:conMpa}
\end{figure*}

Next, we consider the effects of weak lensing distortion of galaxies, which include convergence $\kappa$ and shear $\gamma = \gamma_1 + i\gamma_2$. The effects can be mimicked in simulations by following the paths of photon rays as they traverse the matter field. In practice, the lensing maps can be obtained as follows: first, divide the dark matter particles in a light cone into discrete slices, then trace a bucket of light rays from the observer to the `source' slice along past light cones, and computes the deflection statistics. 

We use the LensTools\footnote{\url{https://github.com/apetri/LensTools}} \citep{lenstools}, a Python computing package to perform the aforementioned multi-lens-plane algorithm \citep{multi-lens} to generate synthetic lensing maps. Note that the resulting maps are on an observer grid, so we first shift the galaxies to their apparent positions via the lensing deflections and then assign the lensing quantities to each galaxy according to its redshift and apparent positions. Since using thin lens planes ($\textless 60$ $h^{-1}$Mpc) may suppress the power spectrum over a broad range of scales  \citep{2020AJ....159..284Z}, we fix the thickness of each lens plane at 100 $h^{-1}$Mpc. 

We derive the tomographic shear maps by projecting the shear values for each galaxy onto a 2D Cartesian grid as
\begin{equation}
    \bm{\hat{\gamma}_{p}}(\bm{\theta}) = \frac{\Sigma_{i\in p}w_{i}\bm{\hat{\gamma}_{i}}(\bm{\theta})}{\Sigma_{i\in p}w_{i}},
\end{equation}
where the index $i$ runs over galaxies in the catalog, index $p$ runs over pixels in the map, $\bm{\hat{\gamma}_{i}} = (\hat{\gamma}_{1,i},\hat{\gamma}_{2,i})$ is the galaxy shear, and $w_{i}$ is the weight and we set $w_{i} = 1$ for simplicity. 

We convert the shear field into a lensing convergence map by performing the Kaiser-Squires inversion (KS-inversion) \citep{KS},
\begin{equation}
    \tilde{\kappa}(\bm{\ell}) = \left (\frac{\ell_{1}^{2} - \ell_{2}^{2}}{\ell_{1}^{2}+\ell_{2}^{2}} \right ) \tilde{\gamma}_{1}(\bm{\ell}) + 2 \left ( \frac{\ell_1 \ell_2}{\ell_{1}^{2}+\ell_{2}^{2}} \right ) \tilde{\gamma}_{2}(\bm{\ell}),
\end{equation}
where $\bm{\ell} = (\ell_1,\ell_2)$ is the multipoles in Fourier space and $\bm{\theta}$ is the angular coordinates of the planes orthogonal to the LOS, and $\tilde{\kappa}$ and $\bm{\tilde{\gamma}} = (\tilde{\gamma_1},\tilde{\gamma_2})$ are the Fourier transforms of $\kappa$ and $\bm{\gamma}$, respectively. After applying the aforementioned operation on the shear field, we can get the convergence field by inverting back to real space\footnote{We use bin2d, ks93 functions from lenspack to perform KS-inversion. \url{https://github.com/CosmoStat/lenspack}}. Note that if the field has gaps due to small pixel size or has missing data due to survey mask, it may create undesirable artifacts in the KS inversion \citep{masked}. To address this issue, one can perform inpainting method \citep{inpainting1,inpainting2} to fill the gaps or masked regions by extrapolating the information from the observed data using the sparsity property of the discrete cosine transform before proceeding with the KS. In our shear map construction, we do not consider mask regions for simplicity, and to ensure we have at least one galaxy per pixel, each map has a size of 512$^2$ pixels, corresponding to a pixel size $\theta_{\text{pix}} \approx 1.17$ arcmin.

After we obtain the desired lensing convergence maps, we add the shape noise as a Gaussian random field with a zero mean on pixels, and its variance is given by \citep{randsig}
\begin{equation}
    \sigma_{\text{rad,pix}}^2 = \frac{\sigma_{\gamma}^2}{\bar{n}_{g}\theta_{\text{pix}}^2},
\end{equation}
where $\sigma_{\gamma}^2$ is the shear variance per component caused by intrinsic ellipticity and measurement error, taken to be $\sigma_{\gamma}^2 = 0.04$, $\bar{n}_{g}$ is the average galaxy number density in a given tomographic bin per steradian, and $\theta_{\text{pix}}$ is the pixel size of the convergence map. 

On the other hand, the tomographic galaxy over-density map can be obtained by
\begin{equation}
    \delta_{p} = \frac{N_{p}\Sigma_{p^{\prime}}\text{v}^{\prime}}{\text{v}_{p}\Sigma_{p^{\prime}}N_{p^{\prime}}} -1,
\end{equation}
where $N_{p} = \Sigma_{i\in p}w_{i}$ is the number of galaxies at a given pixel $p$, and $\text{v}_{p}$ is the effective fraction of the area at pixel $p$. Note that the discrete nature of galaxies introduces a shot-noise contribution to the auto power spectrum, and we model this shot noise and the shape noise of cosmic shear analytically in the next section. 
In Figure \ref{fig:conMpa}, we present the simulated galaxy density map and weak lensing maps of the first and second tomographic bins in 100 deg$^2$.

\section{Power Spectrum}\label{pk}
\subsection{Angular power spectra measurements}

\begin{figure*}
    \centering
    \includegraphics[width=1\linewidth]{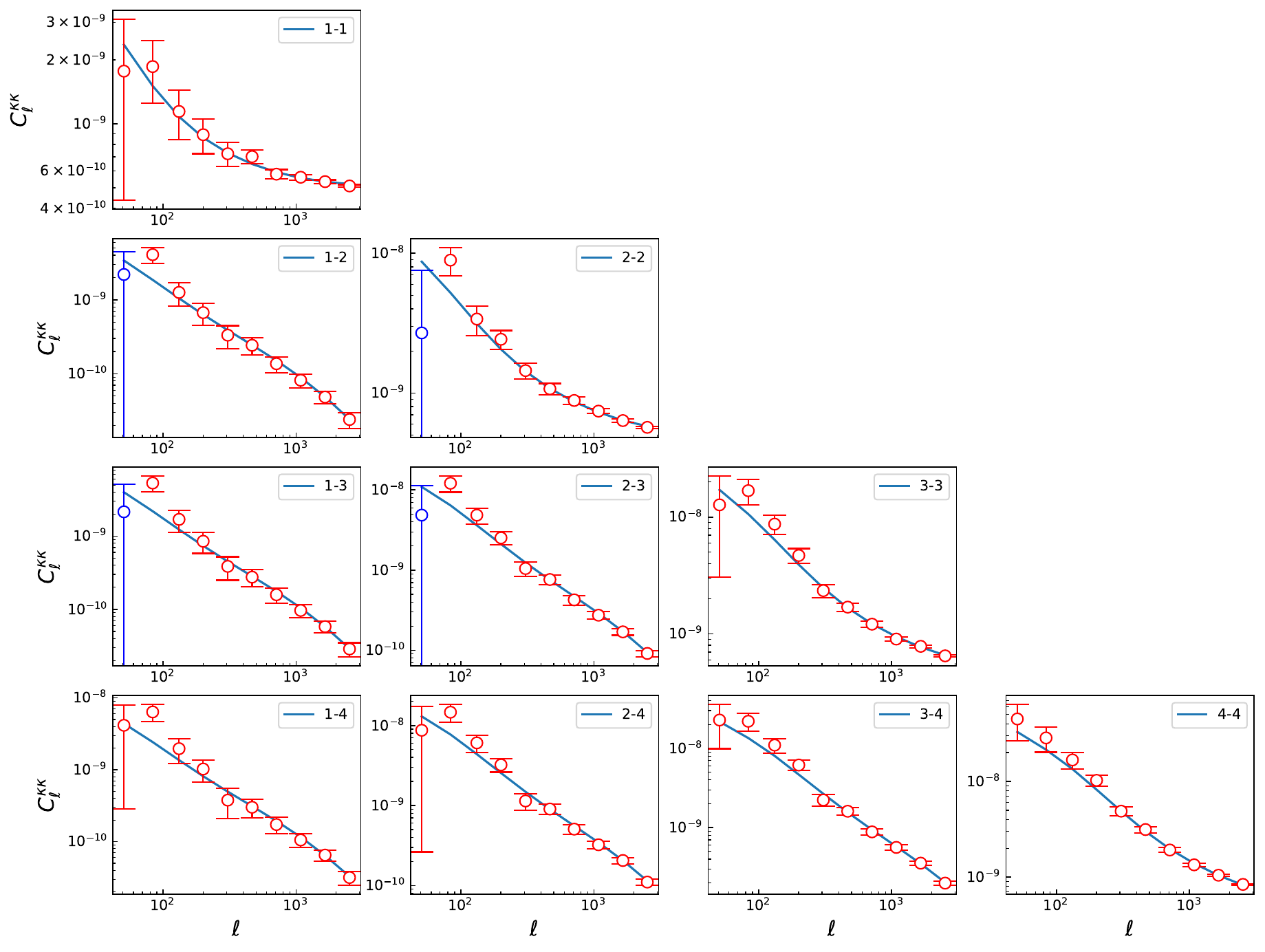}
    \caption{The mock CSST auto and cross shear signal power spectra for the four tomographic bins in 100 deg$^2$. The solid curves show the results of the best-fitting theoretical model, and the data points are the mock data. The blue data points denote the SNR $<1$ which are excluded in the constraint process. To avoid the non-linear effect, we only consider the data points at $\ell < 3000$ in the cosmological analysis. }
    \label{fig:kk}
\end{figure*}

\begin{figure*}
    \centering
    \includegraphics[width=1\linewidth]{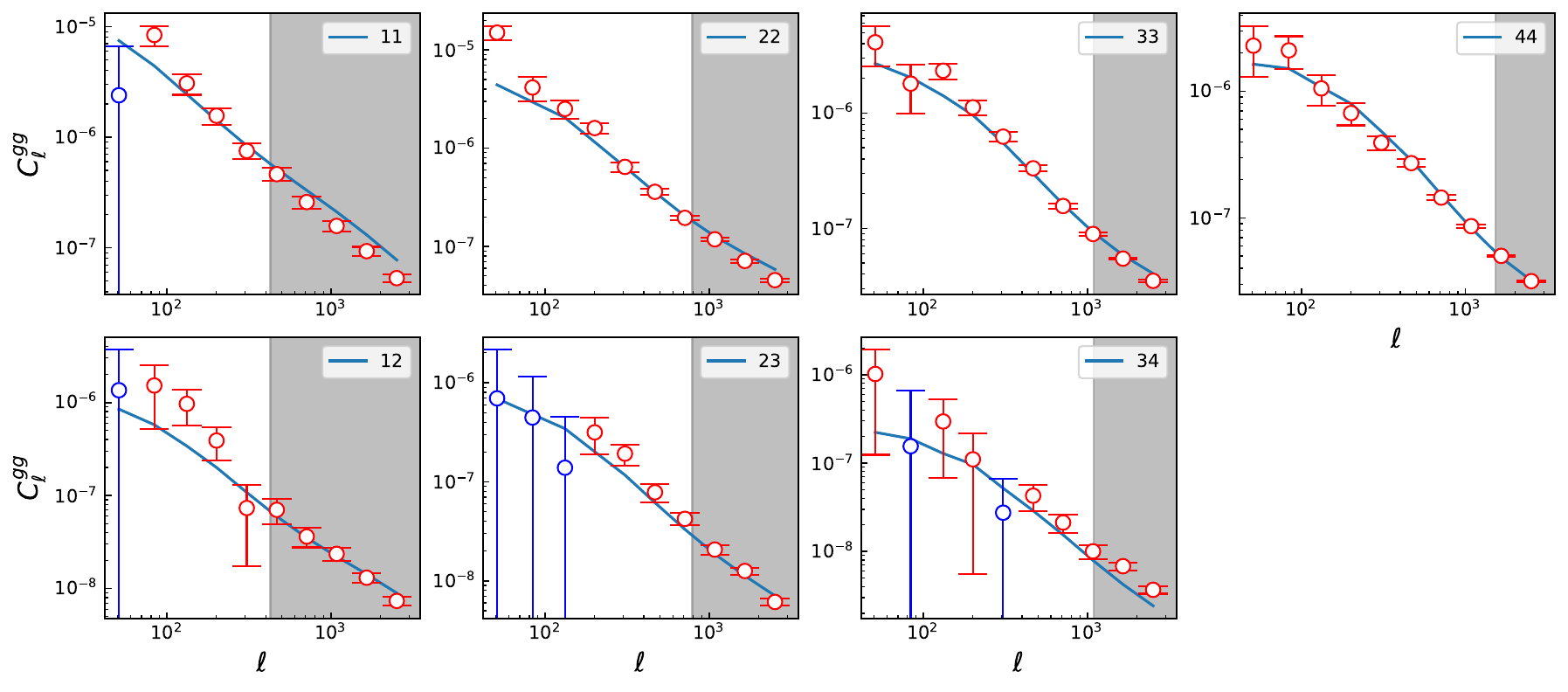}
    \caption{The mock CSST angular galaxy power spectra of the four tomographic bins in 100 deg$^2$. Since the cross-power spectra between nonadjacent bins are quite small, we only consider three cross-power spectra between adjacent bins in our constrain process. The gray regions show the scales that excluded due to the non-linear effects.}
    \label{fig:gg}
\end{figure*}

\begin{figure*}
    \centering
    \includegraphics[width=1\linewidth]{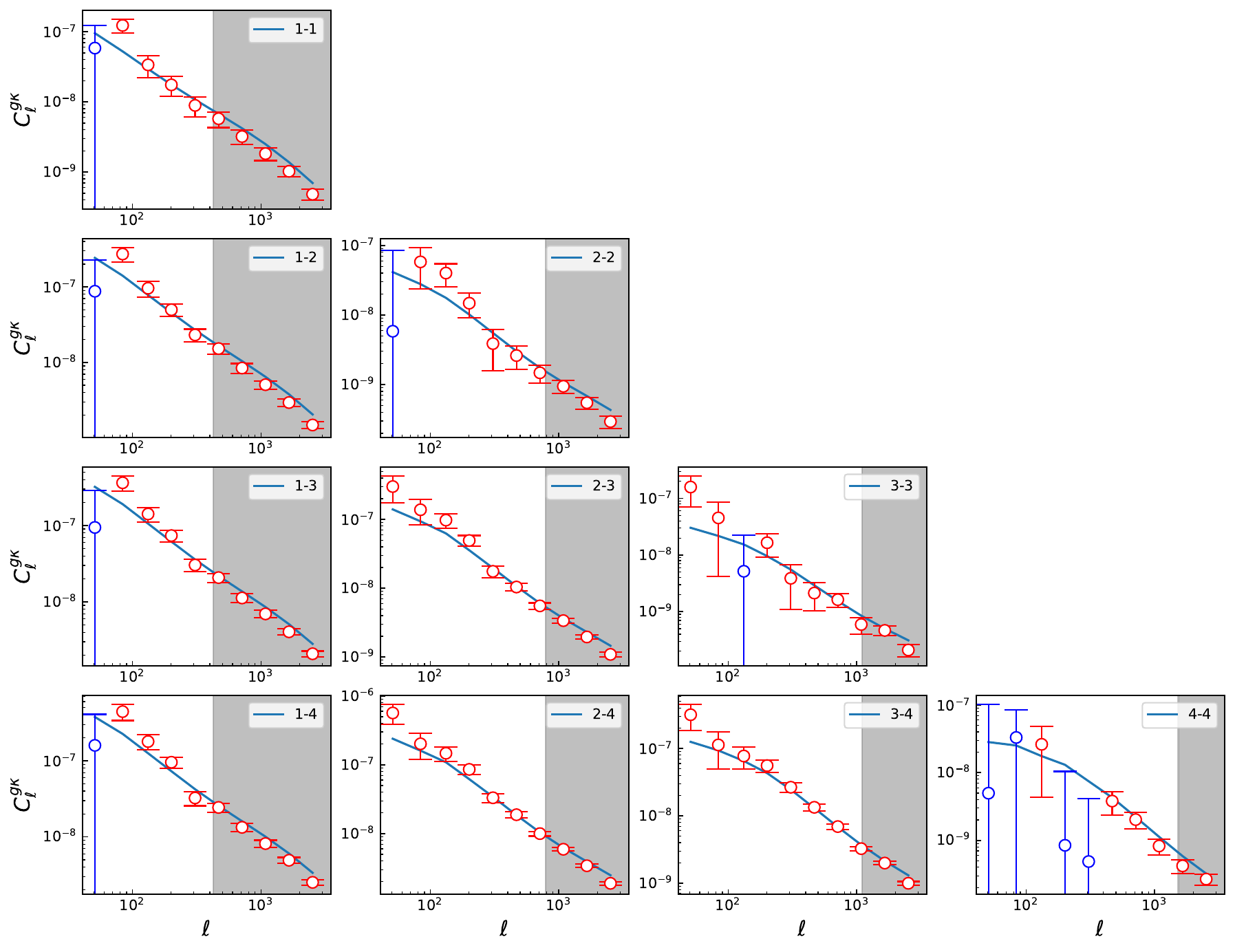}
    \caption{The mock CSST galaxy-galaxy lensing power spectra of the four tomographic bins in 100 deg$^2$. We discard the cross-power spectra with low amplitudes and only consider the galaxy-galaxy lensing power spectrum for the case $a\leq i$ in the analysis, where $a$ and $i$ denote the tomographic bins of the lens and sources samples, respectively. }
    \label{fig:gk}
\end{figure*}

We estimate the angular power spectrum of cosmic shear, galaxy clustering, and galaxy-galaxy lensing using Powerbox \citep{pbox} at $\ell_{\rm max} < 3000$. The results are shown in Figure \ref{fig:kk}, Figure \ref{fig:gg}, and Figure \ref{fig:gk}, along with the theory prediction calculated at the $w$CDM best-fit values using the public CCL \citep{CCL} code\footnote{\url{https://github.com/LSSTDESC/CCL}}.

In order to avoid the non-linear effect, and improve the prediction accuracy of the theoretical model, we only consider the multipoles that satisfy $\ell \leq k_{\text{max}}\chi(\left< z^i \right>)$, where $\left< z^i \right> = \left( \int zn^{i}(z)dz \right) / \left( \int n^{i}(z)dz \right)$ is the mean redshift of the $i$th tomographic bin. We set $k_{\text{max}} = 0.3\, $Mpc$^{-1}$ in the CSST photometric galaxy clustering and galaxy-galaxy lensing analysis \citep{lin}, and the corresponding multipole scale cuts in the four tomographic bins are $\ell_{\text{max}}$ = \{425,789,1097,1523\}. 

\subsection{Theoretical Modeling} \label{sec:tm}
For the shear power spectrum, considering intrinsic alignments and systematics, the measured shear power spectrum for the $i$th and $j$th tomographic bins can be expressed as \citep{2006MNRAS.366..101H}
\begin{equation}
    \tilde{C}_{\kappa\kappa}^{ij}(\ell) = (1+m_{i})(1+m_{j})C_{\kappa\kappa}^{ij}(\ell) + \delta_{ij}^{K}\frac{\sigma_{\gamma}^{2}}{\bar{n}_{g}^{i}},
\end{equation}
where $m_{i}$ is the shear calibration bias or multiplicative error, and the second term accounts for the contribution from the shape noise, $\delta_{ij}^{K}$ is the Kronecker delta function, and $\sigma_{\gamma}^2$ is the shear variance per component, taken to be $\sigma_{\gamma}^2 = 0.04$. For simplicity, we ignore the additive error in the fitting process and assume it can be well controlled in the Stage~IV weak lensing surveys \citep{2012MNRAS.423.3163K,massey2013origins}. The $C_{\kappa\kappa}^{ij}(\ell)$ is the shear signal power spectrum, which is composed of three components \citep{2017MNRAS.465.1454H,2018MNRAS.474.4894J}
\begin{equation}
    C_{\kappa\kappa}^{ij}(\ell) = P_{\kappa}^{ij}(\ell) + C_{\rm II}^{ij}(\ell) + C_{\rm GI}^{ij}(\ell),
\end{equation}
where $P_{\kappa}^{ij}(\ell)$ is the convergence power spectrum, which is the desired cosmic shear power spectrum for cosmological analysis.  $C_{\rm II}^{ij}(\ell)$ and $C_{\rm GI}^{ij}(\ell)$ are the Intrinsic-Intrinsic and Gravitational-Intrinsic power spectrum, accounting for the intrinsic galaxy alignment effects. The former represents the correlation of the intrinsic shape between neighboring galaxies, and the latter denotes the correlation between the intrinsic shapes of foreground galaxies and the cosmic shear of background galaxies \citep{2015SSRv..193....1J}. Under the flat sky assumption and Limber approximation \citep{Limber}, the convergence power spectrum can be obtained by
\begin{equation}
    P_{\kappa}^{ij}(\ell) = \int d\chi \frac{q_{\kappa}^{i}(\chi)q_{\kappa}^{j}(\chi)}{\chi^2}P_{m}\left( \frac{\ell+1/2}{\chi},\chi \right),
\end{equation}
where $\chi$ is the comoving radial distance, $P_{m}$ is the matter power spectrum computed using {\tt CAMB} \citep{camb} and {\tt halofit} code \citep{Taka}, $q_{\kappa}^{i}(\chi)$ is the lensing weighting function of the $i$th tomographic bin
\begin{equation} \label{equ:k}
    q_{\kappa}^{i}(\chi) = \frac{3H_{0}^{2}\Omega_{m}}{2c^2}\frac{\chi}{a(\chi)}\int_{\chi}^{\infty} d\chi^{\prime}n^{i}(z(\chi^{\prime}))\frac{dz}{d\chi^{\prime}}\frac{\chi^{\prime}-\chi}{\chi^{\prime}},
\end{equation}
where $H_0$ is the Hubble constant, $\Omega_m$ is the matter density, $a(\chi)$ is the scale factor corresponding to the comoving distance $\chi$ and $c$ is the speed of light. $n^{i}$ is the normalized redshift distribution of the lens and source galaxies in the tomographic bin $i$.  When modeling the redshift distribution $n^i$, to account for the possible uncertainty in the galaxy redshift distribution, we introduce shift parameters $\Delta z^{i}$ and stretch parameters $\sigma^{i}_{z}$ as free parameters, and we have \citep{redun1,redun2}
\begin{equation}
    n^{i}(z) \rightarrow \frac{n^{i}}{\sigma^{i}_{z}} \left( \frac{z-\left< z^i \right> - \Delta z^{i}}{\sigma_{z}^{i}} + \left< z^i \right> \right),
\end{equation}
where $\left< z^i \right>$ is the mean redshift of the $i$th tomographic bin. The Intrinsic-Intrinsic and Gravitational-Intrinsic power spectrum are given by
\begin{equation}
    C_{\text{II}}^{ij}(\ell) = \int d\chi \frac{F^{i}(\chi)F^{j}(\chi)}{\chi^2} P_{m} \left( \frac{\ell+1/2}{\chi},\chi \right),
\end{equation}
and
\begin{equation}
    C_{\text{GI}}^{ij}(\ell) = \int d\chi \frac{[q_{\kappa}^{i}(\chi)F^{j}(\chi)+q_{\kappa}^{j}(\chi)F^{i}(\chi)]}{\chi^2} P_{m} \left( \frac{\ell+1/2}{\chi},\chi \right),
\end{equation}
Here the weighting function $F^{i}(\chi)$ can be written as \citep{2017MNRAS.465.1454H}
\begin{equation}
    F^{i}(\chi) = -A_{\text{IA}}C_{1}\rho_{c}\frac{\Omega_{m}}{D(\chi)}n^{i}(z(\chi))\frac{dz}{d\chi}\left( \frac{1+z}{1+z_{0}} \right)^{\eta_{\rm IA}} \left( \frac{L_{i}}{L_{0}} \right)^{\beta_{\rm IA}},
\end{equation}
where $A_{\rm IA}$ is the intrinsic alignments amplitude, $C_{1} = 5\times10^{-14}h^{-2}M_{\odot}^{-1}$Mpc$^{3}$, $\rho_{c}$ is the present critical density, $D(\chi)$ is the linear growth factor normalized to unity at $z=0$, $z_{0}=0.6$ and $L_{0}$ are pivot redshift and luminosity. $\eta_{\rm IA}$ and $\beta_{\rm IA}$ represent the relations of redshift and luminosity, respectively. For simplicity, we do not consider luminosity dependence here by fixing $\beta_{\rm IA}=0$, and treat $A_{\rm IA}$ and $\eta_{\rm IA}$ as free parameters in this model.

We then consider the galaxy clustering angular power spectrum for the $i$th and $j$th photo-z bins, which can be calculated from the three-dimensional matter power spectrum under the flat sky assumption and Limber approximation \citep{Limber,hu}
\begin{equation}
    C_{gg}^{ij}(\ell) = \int d\chi \frac{q_{\delta_g}^{i}(\chi) q_{\delta_g}^{j}(\chi)}{\chi^2}P_{m} \left( \frac{\ell + 1/2}{\chi},\chi \right),
\end{equation}
where $q_{\delta_{g}}^{i}(\chi)$ is the weighting function of the $i$th tomographic bin
\begin{equation} \label{equ:g}
    q_{\delta_g}^i(\chi) = b^i[k,z(\chi)]\,n^{i}[z(\chi)]\,\frac{dz}{d\chi},
\end{equation}
where  $b^i[k,z(\chi)]$ is the galaxy bias depending on the scale and redshift. For simplicity, we consider a linear bias model where $b^i(k,z(\chi)) = b^i$ is a constant free parameter for each tomographic bin $i$. Considering the shot-noise term, the total galaxy clustering angular power spectrum can be written as 
\begin{equation}
    \tilde{C}_{gg}^{ij}(\ell) = C_{gg}^{ij}(\ell) + \frac{\delta_{ij}^{K}}{\bar{n}_{g}^{i}},
\end{equation}
where $\bar{n}_{g}^{i}$ is the average angular number density of galaxies in the $i$th bin per steradian.

The angular galaxy-galaxy lensing power spectrum is the cross-correlation between the galaxy clustering and weak lensing signals, it can be expressed as \citep{2018PhRvD..98d3526A}
\begin{equation}
    C_{g\gamma}^{ai}(\ell) = C_{g\kappa}^{ai}(\ell) + C_{g\text{I}}^{ai}(\ell),
\end{equation}
where $C_{g\kappa}$ and $C_{g\text{I}}$ are given by
\begin{equation}
    C_{g\kappa}^{ai} = \int d\chi \frac{q_{\delta_{g}}^{a}(\chi)q_{\kappa}^{i}(\chi)}{\chi^2} P_{m} \left( \frac{\ell+1/2}{\chi},\chi \right),
\end{equation}
and 
\begin{equation}
    C_{g\text{I}}^{ai} = \int d\chi \frac{q_{\delta_{g}}^{a}(\chi)F^{j}(\chi)}{\chi^2} P_{m} \left( \frac{\ell+1/2}{\chi},\chi \right).
\end{equation}

The Equation~ (\ref{equ:k}) and (\ref{equ:g}) indicate that the lensing kernel of high redshift has a rather large overlap with the galaxy clustering kernel of low redshift. On the contrary, the low-redshift lensing kernel has less overlap with the high-redshift galaxy clustering kernel. Results in \cite{lin} have shown significant signal of $C_{g\gamma}^{ai}$ when $a<i$, and a very low amplitude of $C_{g\gamma}^{ai}$ for $a>i$. It is consistent with the physics instincts that the weak lensing signal is the accumulative effect along the LOS and we expect a strong correlation between the background cosmic shear and the foreground galaxy clustering. Since we only simulate the mock data in 100 deg$^2$, the statistical significance of $C_{g\gamma}^{ai}$ for $a>i$ is low, and we only consider the galaxy-galaxy lensing power spectra for the case $a\leq i$ to perform the cosmological analysis as shown in Figure~\ref{fig:gk}. For the full CSST survey with 17500 deg$^2$, we need to include all cross-power spectra between different tomographic bins.

\subsection{Covariance computation}

\begin{figure}
    \centering
    \includegraphics[width=1\linewidth]{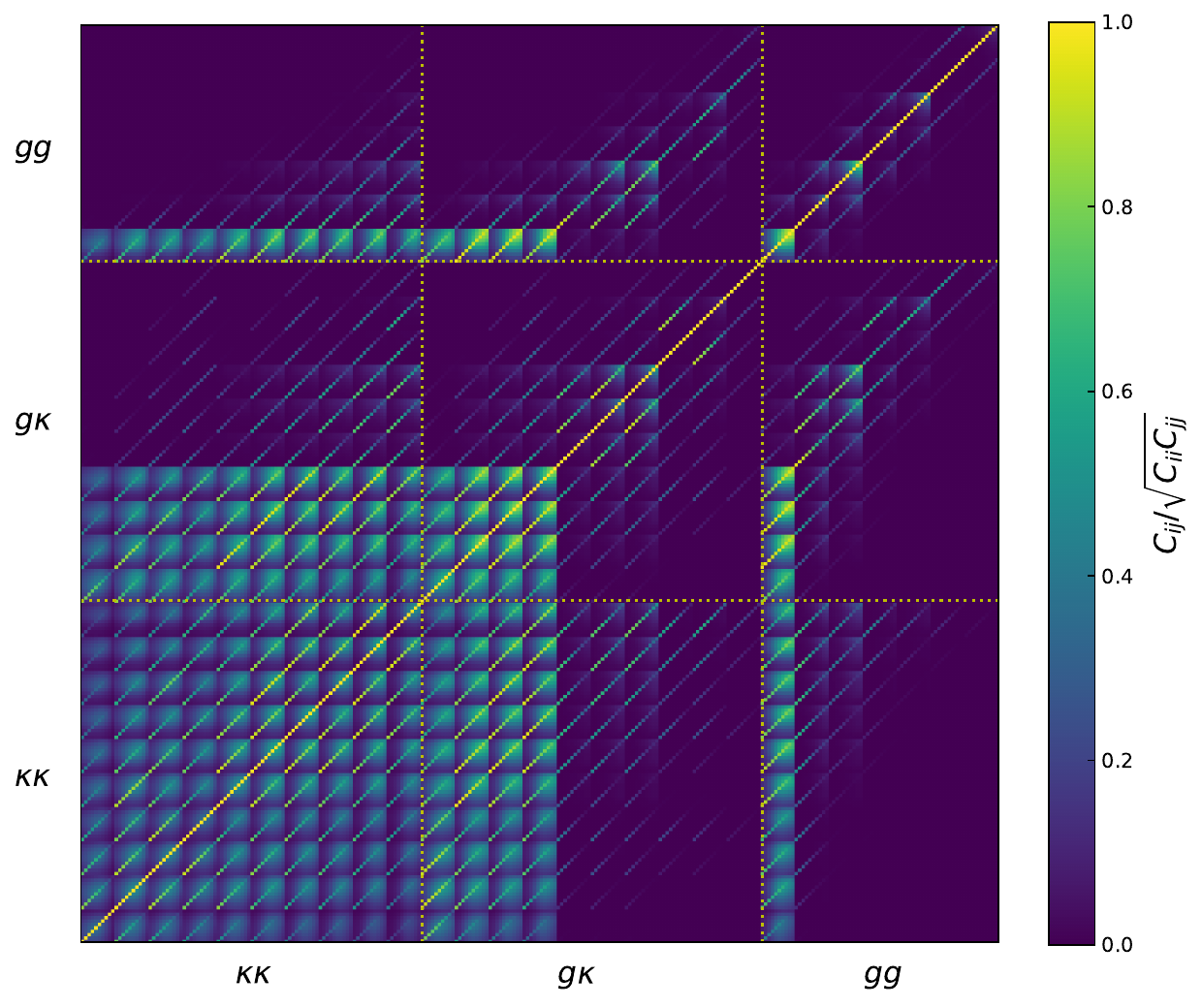}
    \caption{The normalized full covariance matrix (or the correlation coefficient, $C_{ij}/\sqrt{C_{ii}C_{jj}}$, where $C_{ij}$ is the elements of the covariance matrix), for the joint data vector of galaxy clustering, galaxy-galaxy lensing, and cosmic shear including non-Gaussian terms. }
    \label{fig:cov}
\end{figure}



The joint covariance of cosmic shear, galaxy clustering, and galaxy-galaxy lensing power spectrum can be analytically decomposed into Gaussian and non-Gaussian components. The Gaussian term can be expressed as a four-point function, which can be estimated as
\begin{equation}
\label{cov}
\begin{split}
    &\text{Cov}_{ABA^{\prime}B^{\prime}}^{ijkl}(\ell) \equiv \text{Cov} \left[ \tilde{C}_{AB}^{ij}(\ell) \tilde{C}_{A^{\prime}B^{\prime}}^{kl}(\ell^{\prime}) \right]\\
    &= \frac{\delta_{\ell \ell^{\prime}}^{K}}{(2\ell+1)f_{\rm sky}\Delta \ell} \left[ \tilde{C}_{AA^{\prime}}^{ik}(\ell) \tilde{C}_{BB^{\prime}}^{jl}(\ell^{\prime}) + \tilde{C}_{AB^{\prime}}^{il}(\ell) \tilde{C}_{BA^{\prime}}^{jk}(\ell^{\prime}) \right],
\end{split}
\end{equation}
where $A, A^{\prime}, B, B^{\prime}$ $\in$ \{g,$\kappa$\} denote different type of tracers, and $f_{\text{sky}}$ is the sky coverage fraction corresponding to 100 deg$^2$. And we use $N_{\ell} = 10$ logarithm-spaced multipole bins at $\ell<3000$.

On the other hand, the non-Gaussian contribution comprises two components: the connected four-point covariance (cNG), arising from the non-Gaussian distribution of the matter field due to late-time non-linear evolution \citep{cNG}, and the super-sample covariance (SSC), which accounts for the correlations between Fourier modes used in analysis and super-survey modes \citep{SSC}. The computation of non-Gaussian components utilizes the implementation provided by CCL code \citep{CCL}. In the calculation, we adopt the NFW profile \citep{NFW}, along with the concentration-mass relation \citep{c-m}, halo mass function \citep{tink08} and halo bias \citep{tink10}. We simplify the treatment of partial sky coverage for the non-Gaussian contribution by scaling it with the light-cone sky coverage $f_{\rm sky}$. More details can be found in \cite{yao}. Figure \ref{fig:cov} shows the normalized full covariance matrix for the joint data vector of galaxy clustering, galaxy-galaxy lensing, and cosmic shear with non-Gaussian terms, which introduce the off-diagonal features.

We also use the jackknife method combined with pseudo-$C_{\ell}$ or MASTER method to estimate the sample covariance from the data \citep{1973ApJ...185..413P,pseudo2,pseudo3}, which is implemented in NAMASTER code\footnote{\url{https://github.com/LSSTDESC/NaM ster}}\citep{2019MNRAS.484.4127A}. In order to ensure the obtained covariance matrix is non-singular and can get the desired precision matrix \citep{2024arXiv240213783L}, we need to divide the samples of each tomographic bin into more than 250 patches, but we find the region of patches are not big enough for the method to work well. Therefore we choose to use the analytical modeling covariance in our constrain process.


\section{CONSTRAINT AND RESULT}\label{fitting}
\subsection{Fitting Method}
The $\chi^2$ method is adopted to fit the mock data of the CSST photometric surveys, and the likelihood function $\mathcal{L}\propto {\rm exp}(-1/2\,\chi^2)$ is given by
\begin{equation}
    \text{ln} \mathcal{L}(\bm{D}|\bm{p}) \propto -\frac{1}{2}[\bm{D} - \bm{M}(\bm{p})]^{T}\textbf{Cov}^{-1}[\bm{D} - \bm{M}(\bm{p})] ,
\end{equation}
where $\bm{D}$ is the data vector of the angular galaxy, galaxy-galaxy lensing, and cosmic shear power spectra, $\bm{M}$ is the corresponding theoretical power spectra, and $\textbf{Cov}$ is the joint covariance matrix. 

\renewcommand{\arraystretch}{1.2}
\setlength{\tabcolsep}{10pt}
\begin{table}[t]
    \centering
    \caption{The free parameters considered in our constraint process. The first column shows the names of the free parameters, and the second and third columns show the fiducial values and the priors of the parameters, respectively. Uniform priors are described by $\mathcal{U}(x,y)$, with $x$ and $y$ denoting the prior range. The Gaussian priors are represented by $\mathcal{N}$$(\sigma,\mu)$, and $\sigma$ and $\mu$ are the mean and standard deviation, respectively.}
    \begin{tabular}{ccc}
    \hline
    \hline
     Parameter & Fiducial Value & Prior  \\
    \hline
      & \textbf{Cosmology} \\
     $h$ & 0.6766 & $\mathcal{U}(0.4,1.0)$ \\
     $\Omega_{m}$ & 0.3111 & $\mathcal{U}(0.1, 0.6)$    \\
     $\Omega_{b}$ & 0.0490 & $\mathcal{U}(0.01, 0.09)$ \\
     $\sigma_{8}$ & 0.8102 & $\mathcal{U}(0.4, 1.2)$ \\ 
     $n_{s}$ & 0.9665 & $\mathcal{U}(0.7, 1.2)$\\
     $w$ & -1 & $\mathcal{U}$(-1.8, 0.2) \\
     \hline
     & \textbf{Intrinsic alignment} \\
     $A_{\text{IA}}$ & 0 & $\mathcal{U}$(-5, 5) \\
     $\eta_{\text{IA}}$ & 0 & $\mathcal{U}$(-5, 5) \\
     \hline
     & \textbf{Galaxy bias} \\
     $b_{g}^{i}$ & - & $\mathcal{U}$(0, 5) \\
     \hline
     & \textbf{Photo-z shift} \\
     $\Delta z^{i}$ & (0, 0, 0, 0) & $\mathcal{N}$(0, 0.01) \\
     \hline
     & \textbf{Photo-z stretch} \\
     $\sigma_{z}^{i}$ & (1, 1, 1, 1) & $\mathcal{N}$(1, 0.05) \\
     \hline
     & \textbf{Shear calibration} \\
     $m_{i}$ & (0, 0, 0, 0) & $\mathcal{N}$(0, 0.01) \\ 
     \hline
     \hline
    \end{tabular}
    \label{tab:fit}
\end{table}

\begin{figure*}[t]
    \centering
    \includegraphics[width=0.95\linewidth]{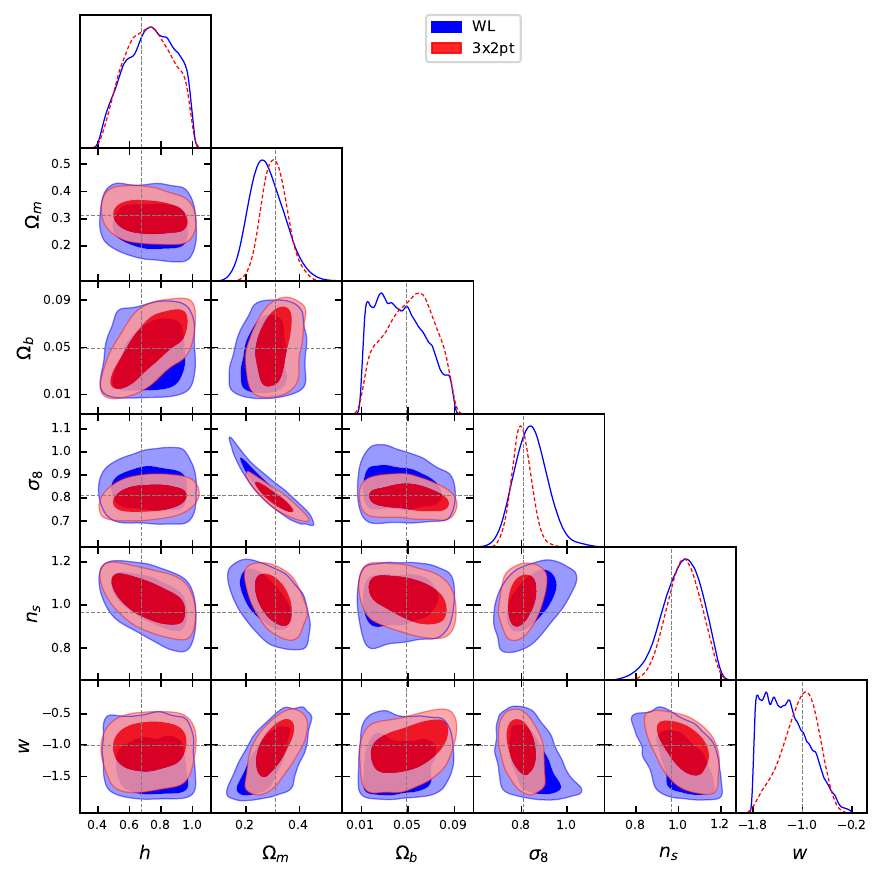}
    \caption{The constraint results of the six cosmological parameters from the CSST weak lensing only and joint 3$\times$2pt constraints in 100 deg$^2$ survey area. The 1$\sigma$ (68.3\%) and 2$\sigma$ (95.5\%) CLs are shown. The vertical and horizontal dotted lines represent the fiducial values of these parameters.}
    \label{fig:cosmopars}
\end{figure*}

We make use of {\tt emcee}\footnote{\url{https://github.com/dfm/emcee}}\citep{emcee}, which is a stable and well-tested Python implementation of the affine-invariant ensemble sampler of Markov chain Monte Carlo (MCMC) \citep{2013PASP..125..306F}, to perform the fitting process. We set the walkers to be 50 and the number of steps to be 25000 for each walker to ensure the chain reaches the convergence. After the burn-in and thinning processes, we combine all chains and obtain about 10,000 chain points to illustrate the PDFs of the free parameters.  We summarize the free parameters included in the present analysis, and their fiducial values and priors are shown in Table \ref{tab:fit}. We have 6 cosmological parameters and 18 systematic parameters in the fitting process. For cosmological parameters, intrinsic alignment, and galaxy bias, we use a flat prior denoted by $\mathcal{U}$. We employ the Gaussian priors for photo-$z$ and shear calibration parameters presented by $\mathcal{N}$, assuming the CSST photometric survey can provide good photo-$z$ results \citep{cao} and control the relevant systematic effects \citep{gong}.

\subsection{Constraint results}
In Figure \ref{fig:cosmopars}, we show the marginalized contours and one-dimensional (1D) PDFs for the six cosmological parameters from weak lensing only and joint 3$\times$2pt constraint at 68.3\% and 95.5\% confidence levels (CLs). The vertical and horizontal dotted lines show the fiducial values of the parameters. 
For the case of galaxy clustering, since it is limited by the sky area of our simulation, we do not obtain effective constraints on the parameters using it alone and would not present the results here.

\begin{figure}
    \centering
    \includegraphics[width=1\linewidth]{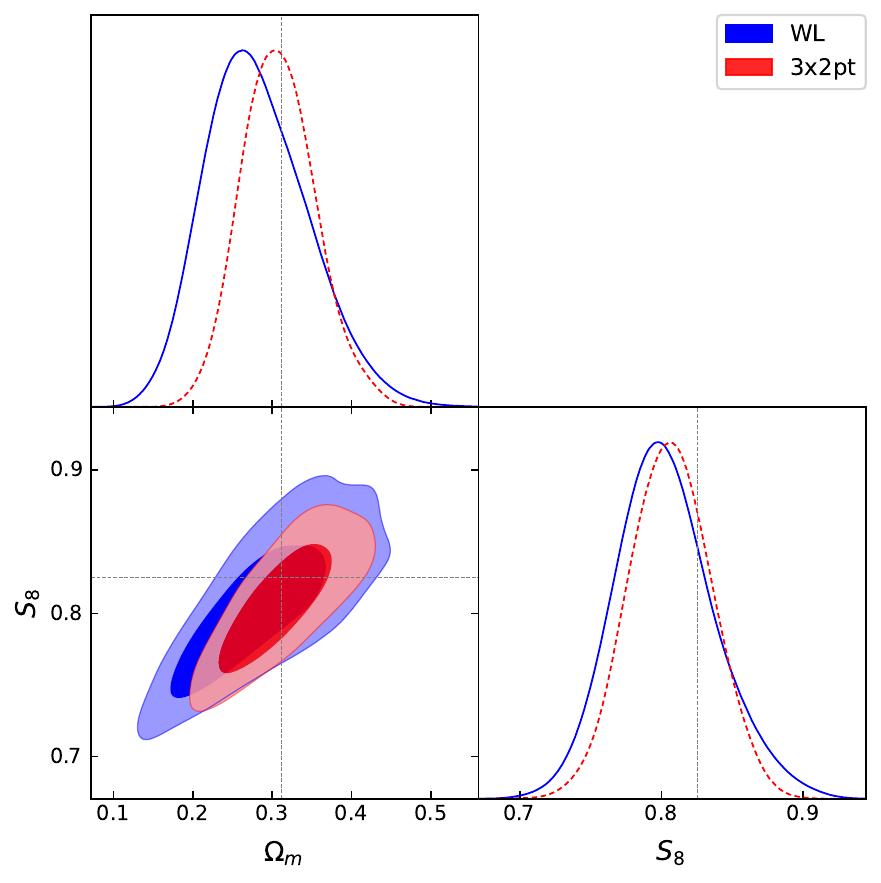}
    \caption{The marginalized contour maps (68.3\% and 95.5\% CLs) of $\Omega_m$ vs. $S_8 \equiv \sigma_8(\Omega_m/0.3)^{0.5}$ for the CSST weak lensing only (blue) and joint 3$\times$2pt (red) constraints in 100 deg$^2$.}
    \label{fig:om-S8}
\end{figure}

\begin{figure}
    \centering
    \includegraphics[width=1\linewidth]{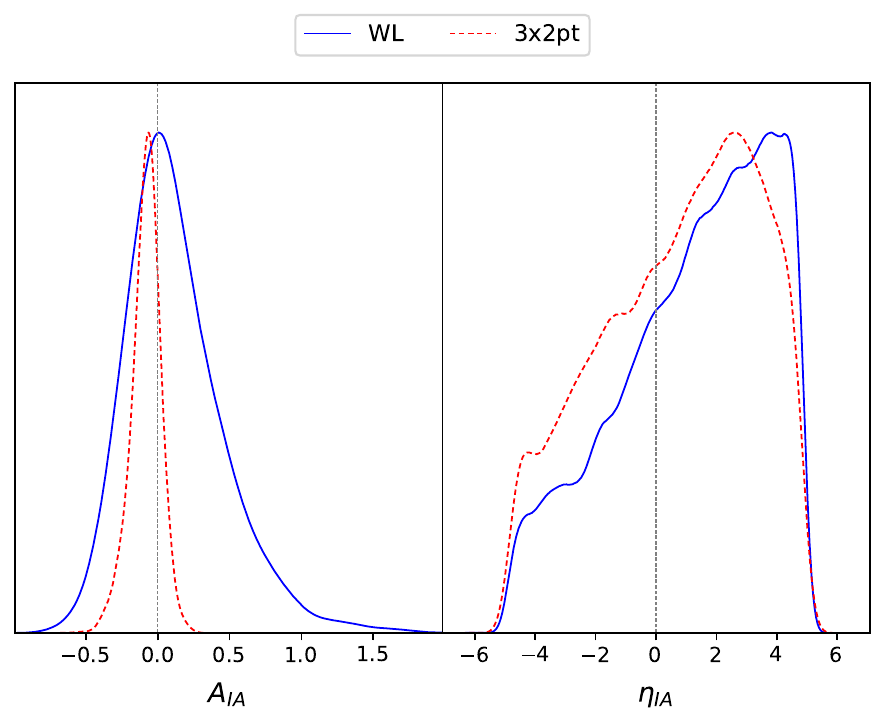}
    \caption{The 1D PDFs of the intrinsic alignment parameters $A_{\rm IA}$ and $\eta_{\rm IA}$. The blue solid and red dashed curves denote the fitting results of the CSST weak lensing only and joint constraint results in 100 deg$^2$, respectively}
    \label{fig:IA}
\end{figure}

\begin{figure*}
    \centering
    \includegraphics[width=0.8\linewidth]{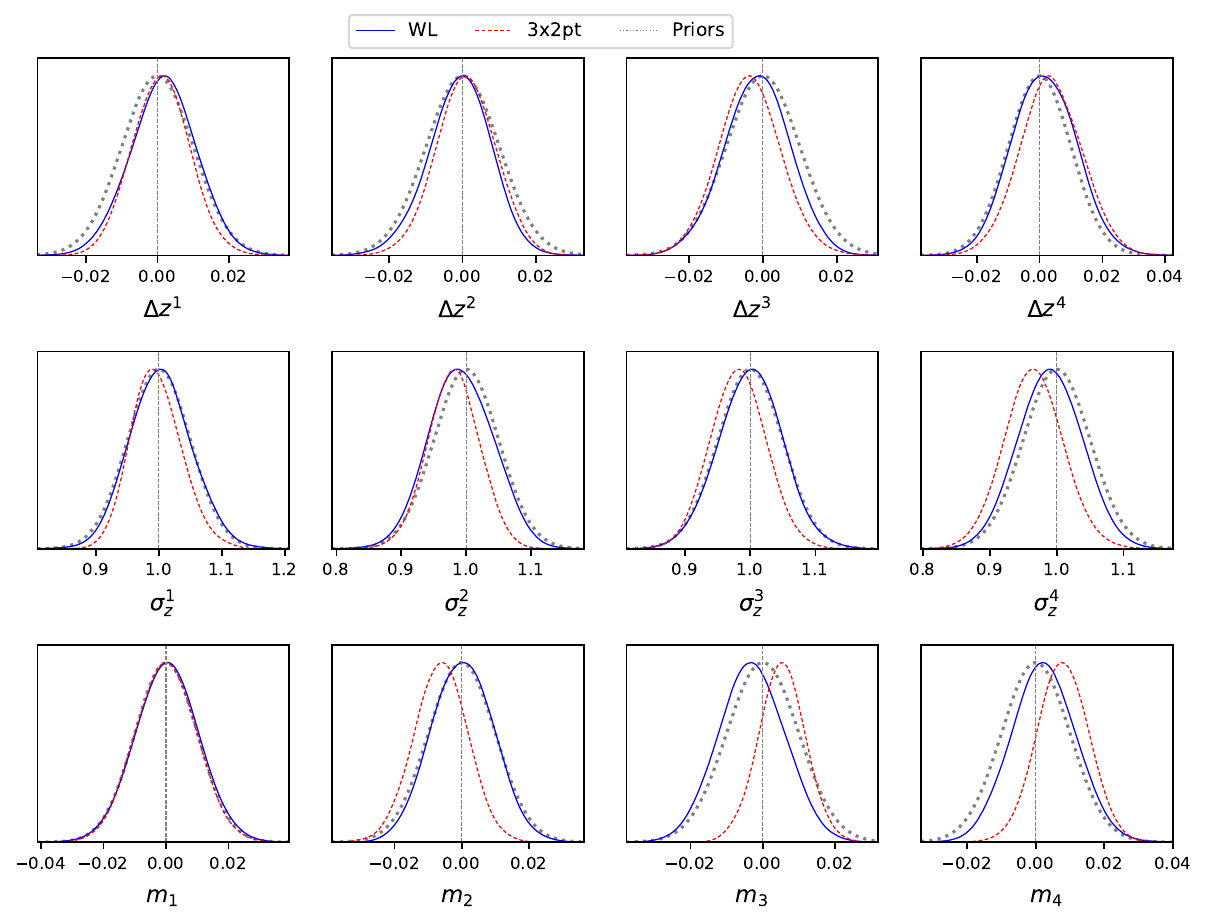}
    \caption{The 1D PDFs of the systematic parameters $\Delta z^{i}$, $\sigma_{z}^{i}$ and $m_{i}$ for the CSST weak lensing only (blue solid curves) and joint constraints (red dotted curves) in 100 deg$^2$ survey area for the four tomographic bins. The gray dotted line represent the Gaussian distribution of priors.}
    \label{fig:z-m}
\end{figure*}

We can see that the cosmological parameters can be correctly and well constrained using the mock data in this 100 deg$^2$ survey area. The fiducial values are within 1$\sigma$ CL of the parameter fitting results for both weak lensing only and joint 3$\times$2pt constraints. We find that, in the current 100 deg$^2$ mock data for the CSST weak lensing survey, we obtain 23\%, 9\%, and 28\% constraint accuracies for $\Omega_m$, $\sigma_8$, and $w$, respectively, with $\Omega_m=0.281^{+0.055}_{-0.074}$, $\sigma_8=0.843^{+0.066}_{-0.078}$, and $w=-1.26^{+0.19}_{-0.51}$. When combining the mock data of galaxy clustering, weak lensing, and galaxy-galaxy lensing, i.e. the joint 3$\times$2pt constraints, we can obtain tighter constraints results for all cosmological parameters, especially for $\Omega_m$, $\sigma_8$, and $w$ (see Figure~\ref{fig:cosmopars}, improved by factors of $\sim1.5-2$). 
We also notice that, as shown in Figure \ref{fig:cosmopars}, the constraints on some parameters are prior dominated in the weak lensing only case (e.g. $w$) and even in the 3$\times$2pt case (e.g. $\Omega_b$ and $h$). This is because that the 100 deg$^2$ survey area used in the analysis is small, and this issue can be significantly reduced in the full CSST survey \citep[e.g.][]{Miao,lin,LinAxion}.

To break the degeneracy between $\Omega_m$ and $\sigma_8$, we also show the constraint results of $\Omega_m$ vs. the less degenerate parameter $S_8 \equiv \sigma_8(\Omega_m/0.3)^{0.5}$ in Figure \ref{fig:om-S8}. We find that $S_8 = 0.80^{+0.03}_{-0.04}$ and $S_8 = 0.81\pm 0.03$ for weak lensing only and joint constraints, with constrain accuracies 9\% and 7\%, respectively. Note that $S_8$ does
not capture the best-constrained direction perpendicular to the $\Omega_m -\sigma_8$ degeneracy, as the scaling  combination with $\Omega_m$ and $\sigma_8$ depends on the weighting of the angular scales entering the analysis \citep{S81}, leading to less improvement for joint 3$\times$2pt constrains.

We check the goodness-of-fit using the reduced chi-square $\chi^2_{\text{red}} \equiv \chi^2 / \text{DoF}$, where DoF is the number of degrees of freedom, defined as the number of data points in the data vector minus the total number of parameters constrained. We find that the total $\chi^2$ values are 86.4 and 283.0 for weak lensing only and joint 3$\times$2pt analysis, with 72 and 168 degrees of freedom (96 data points for weak lensing, 192 data points for joint analysis and 24 free parameters), corresponding to the $\chi^{2}_{\text{red}}=1.2$ and 1.68, respectively.

The constraint result of these cosmological parameters is comparable with DES Y3 galaxy clustering and galaxy-galaxy lensing measurements (2$\times$2pt) in harmonic space \citep{comres1} and in configuration space \citep{comres2}. Note that the current constraint accuracies derived from the mock data are limited by the 100 deg$^2$ simulated sky area of the light cone, and the result could be significantly improved if considering the full CSST photometric survey with $\sim$ 17500 deg$^2$ sky area. Previous theoretical studies indicate that the constraint accuracies of the cosmological parameters may achieve a few percent level or even higher in the full CSST survey \citep[e.g.][]{Miao,lin,LinAxion}.

The 1D PDFs of the intrinsic alignment parameters $A_{\rm IA}$ and $\eta_{\rm IA}$ are shown in Figure \ref{fig:IA} for the weak lensing only and joint surveys in 100 deg$^2$. We can find that there is a significant improvement on the constraint of $A_{\rm IA}$ in the joint analysis, with a factor $\sim$ 2.3 tighter compared to the weak lensing alone. The 1D PDFs of the redshift shift bias $\Delta z^{i}$, stretch factor $\sigma_{z}^{i}$ and multiplicative error $m_i$ are shown in Figure \ref{fig:z-m}. Blue solid lines and red dashed lines show the results from the CSST weak lensing and joint surveys in 100 deg$^2$, respectively. The gray dotted lines represent the distribution of the Gaussian priors. We can see that the constraints on these systematic parameters are relatively weak and prior dominated, due to the current small survey area we use with low statistics. As indicated in the previous theoretical studies, the constraint accuracy of these systematic parameters can be greatly improved in the full CSST survey \citep{gong,Miao,lin,LinAxion}.

\section{SUMMARY} \label{summary}
In this work, we explore the constraints on the cosmological parameters under the $w$CDM model for the CSST photometric surveys in harmonic space. 
We construct the mock galaxy catalog based on JiuTian-1G simulation, and consider the CSST instrumental design and strategy of the CSST photometric survey. We select special viewing angles to build the past light cones to reduce the repeated structures, and perform the multi-lens-plane algorithm to generate the weak lensing catalog in 100 deg$^2$. 

We derive the galaxy redshift distribution from the mock catalog, and divide it into four photometric redshift bins. To account for the photo-$z$ uncertainties in galaxies redshift distribution, we adopt two parameters model including shift and stretch parameters. Then we obtain the tomographic galaxy over-density map and weak lensing convergence map from the mock catalog, and estimate the angular power spectra of galaxy clustering, cosmic shear and galaxy-galaxy lensing. 

After obtaining the mock data, the MCMC technique is employed in the fitting process. The fitting results of the six cosmological parameters under the $w$CDM model are consistent with the fiducial values in 1$\sigma$ CL. For the joint 3$\times$2pt constrains, we obtain tight constraints on the key cosmological parameters in this CSST 100 deg$^2$ mock survey, which is even comparable to the current Stage~III surveys. For the full CSST survey with 17500 deg$^2$, the constraint accuracies can be significantly improved further. These results indicate that the CSST photometric survey is powerful to explore the Universe, and could greatly improve the relevant cosmological studies.


\begin{acknowledgments}
QX and YG acknowledge the support from National Key R\&D Program of China grant Nos. 2022YFF0503404, 2020SKA0110402, and the CAS Project for Young Scientists in Basic Research (No. YSBR-092). XLC acknowledges the support of the National Natural Science Foundation of China through Grant Nos. 11473044 and 11973047, and the Chinese Academy of Science grants ZDKYYQ20200008, QYZDJ-SSW-SLH017, XDB 23040100, and XDA15020200. QG acknowledges the support from the National Natural Science Foundation of China (NSFC No.12033008). AI acknowledges the support of the National SKA Program of China with grant No. 2022SKA0110100 and the Alliance of International Science Organizations Visiting Fellowship on Mega-Science Facilities for Early-Career Scientists (Grant No: ANSO-VF-2022-01, and ANSO-VF-2024-01). The Jiutian simulations were conducted under the support of the science research grants from the China Manned Space Project with NO. CMS-CSST-2021-A03. This work is also supported by science research grants from the China Manned Space Project with Grant Nos. CMS-CSST-2021-B01 and CMS-CSST-2021-A01.
\end{acknowledgments}

%

\vspace{5mm}





\begin{figure}
    \centering
    \includegraphics[width=0.6\linewidth]{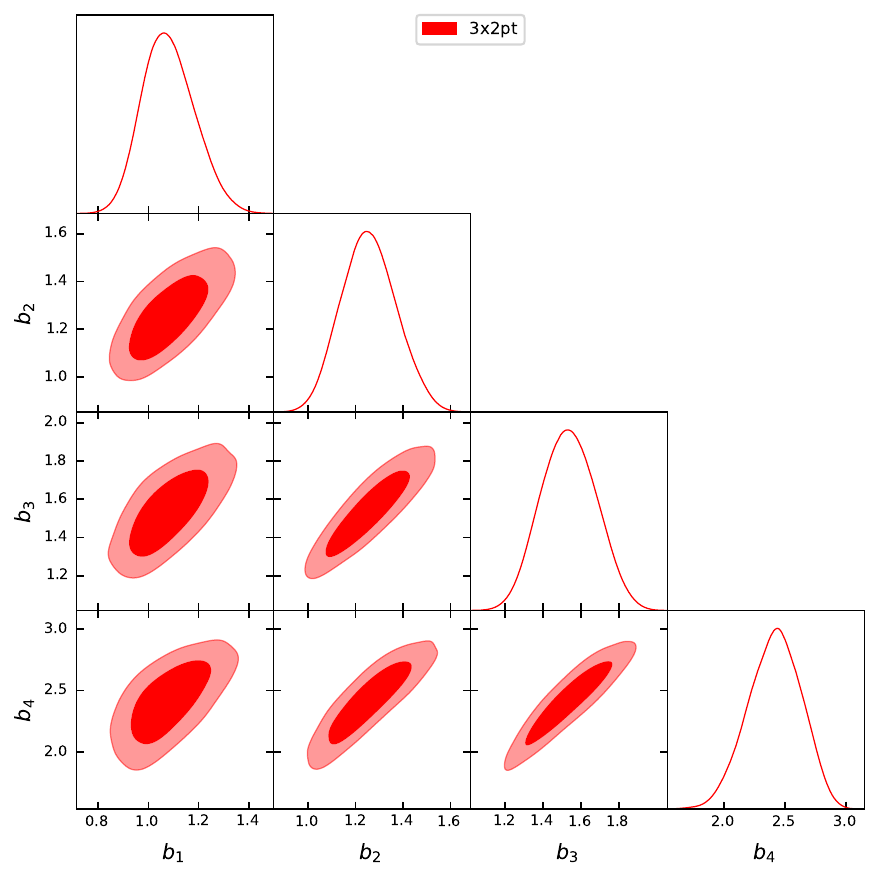}
    \caption{The contour maps (68.3\% and 95.5\% CLs) and 1D PDFs of the linear galaxy bias for the four tomographic bins from the CSST galaxy clustering, weak lensing, and galaxy-galaxy lensing joint surveys in 100 deg$^2$.}
    \label{fig:gbias}
\end{figure}

\appendix
In Figure \ref{fig:gbias}, we show the CSST joint constraint results in 100 deg$^2$ of the linear galaxy bias of the four tomographic bins with the contour maps and 1D PDFs. Note that there are no fiducial values for $b_{g}^i$, since they are not the input parameters in the simulation. The best-fitting values and errors of $b_{i}^{g}$ of the four tomographic bins are \{$1.08^{+0.09}_{-0.12}$, $1.26\pm 0.11$, $1.54\pm 0.14$, $2.42^{+0.23}_{-0.21}$\}. We compare our results with the DES VS data \citep{gbias} and find that for the similar low redshift ranges , our galaxy bias parameters are consistent with those obtained by DES within 1$\sigma$ confidence level.

\bibliography{wl3x2pt}{}

\begin{thebibliography}{}
\expandafter\ifx\csname natexlab\endcsname\relax\def\natexlab#1{#1}\fi
\providecommand{\url}[1]{\href{#1}{#1}}
\providecommand{\dodoi}[1]{doi:~\href{http://doi.org/#1}{\nolinkurl{#1}}}
\providecommand{\doeprint}[1]{\href{http://ascl.net/#1}{\nolinkurl{http://ascl.net/#1}}}
\providecommand{\doarXiv}[1]{\href{https://arxiv.org/abs/#1}{\nolinkurl{https://arxiv.org/abs/#1}}}

\bibitem[{{Abbott} {et~al.}(2018{\natexlab{a}}){Abbott}, {Abdalla}, {Alarcon},
  {Aleksi{\'c}}, {Allam}, {Allen}, {Amara}, {Annis}, {Asorey}, {Avila},
  {Bacon}, {Balbinot}, {Banerji}, {Banik}, {Barkhouse}, {Baumer}, {Baxter},
  {Bechtol}, {Becker}, {Benoit-L{\'e}vy}, {Benson}, {Bernstein}, {Bertin},
  {Blazek}, {Bridle}, {Brooks}, {Brout}, {Buckley-Geer}, {Burke}, {Busha},
  {Campos}, {Capozzi}, {Carnero Rosell}, {Carrasco Kind}, {Carretero},
  {Castander}, {Cawthon}, {Chang}, {Chen}, {Childress}, {Choi}, {Conselice},
  {Crittenden}, {Crocce}, {Cunha}, {D'Andrea}, {da Costa}, {Das}, {Davis},
  {Davis}, {De Vicente}, {DePoy}, {DeRose}, {Desai}, {Diehl}, {Dietrich},
  {Dodelson}, {Doel}, {Drlica-Wagner}, {Eifler}, {Elliott}, {Elsner},
  {Elvin-Poole}, {Estrada}, {Evrard}, {Fang}, {Fernandez}, {Fert{\'e}},
  {Finley}, {Flaugher}, {Fosalba}, {Friedrich}, {Frieman},
  {Garc{\'\i}a-Bellido}, {Garcia-Fernandez}, {Gatti}, {Gaztanaga}, {Gerdes},
  {Giannantonio}, {Gill}, {Glazebrook}, {Goldstein}, {Gruen}, {Gruendl},
  {Gschwend}, {Gutierrez}, {Hamilton}, {Hartley}, {Hinton}, {Honscheid},
  {Hoyle}, {Huterer}, {Jain}, {James}, {Jarvis}, {Jeltema}, {Johnson},
  {Johnson}, {Kacprzak}, {Kent}, {Kim}, {King}, {Kirk}, {Kokron}, {Kovacs},
  {Krause}, {Krawiec}, {Kremin}, {Kuehn}, {Kuhlmann}, {Kuropatkin}, {Lacasa},
  {Lahav}, {Li}, {Liddle}, {Lidman}, {Lima}, {Lin}, {MacCrann}, {Maia},
  {Makler}, {Manera}, {March}, {Marshall}, {Martini}, {McMahon}, {Melchior},
  {Menanteau}, {Miquel}, {Miranda}, {Mudd}, {Muir}, {M{\"o}ller}, {Neilsen},
  {Nichol}, {Nord}, {Nugent}, {Ogando}, {Palmese}, {Peacock}, {Peiris},
  {Peoples}, {Percival}, {Petravick}, {Plazas}, {Porredon}, {Prat}, {Pujol},
  {Rau}, {Refregier}, {Ricker}, {Roe}, {Rollins}, {Romer}, {Roodman},
  {Rosenfeld}, {Ross}, {Rozo}, {Rykoff}, {Sako}, {Salvador}, {Samuroff},
  {S{\'a}nchez}, {Sanchez}, {Santiago}, {Scarpine}, {Schindler}, {Scolnic},
  {Secco}, {Serrano}, {Sevilla-Noarbe}, {Sheldon}, {Smith}, {Smith}, {Smith},
  {Soares-Santos}, {Sobreira}, {Suchyta}, {Tarle}, {Thomas}, {Troxel},
  {Tucker}, {Tucker}, {Uddin}, {Varga}, {Vielzeuf}, {Vikram}, {Vivas},
  {Walker}, {Wang}, {Wechsler}, {Weller}, {Wester}, {Wolf}, {Yanny}, {Yuan},
  {Zenteno}, {Zhang}, {Zhang}, {Zuntz}, \& {Dark Energy Survey
  Collaboration}}]{DESreal1}
{Abbott}, T.~M.~C., {Abdalla}, F.~B., {Alarcon}, A., {et~al.}
  2018{\natexlab{a}}, \prd, 98, 043526, \dodoi{10.1103/PhysRevD.98.043526}

\bibitem[{{Abbott} {et~al.}(2018{\natexlab{b}}){Abbott}, {Abdalla}, {Alarcon},
  {Aleksi{\'c}}, {Allam}, {Allen}, {Amara}, {Annis}, {Asorey}, {Avila},
  {Bacon}, {Balbinot}, {Banerji}, {Banik}, {Barkhouse}, {Baumer}, {Baxter},
  {Bechtol}, {Becker}, {Benoit-L{\'e}vy}, {Benson}, {Bernstein}, {Bertin},
  {Blazek}, {Bridle}, {Brooks}, {Brout}, {Buckley-Geer}, {Burke}, {Busha},
  {Campos}, {Capozzi}, {Carnero Rosell}, {Carrasco Kind}, {Carretero},
  {Castander}, {Cawthon}, {Chang}, {Chen}, {Childress}, {Choi}, {Conselice},
  {Crittenden}, {Crocce}, {Cunha}, {D'Andrea}, {da Costa}, {Das}, {Davis},
  {Davis}, {De Vicente}, {DePoy}, {DeRose}, {Desai}, {Diehl}, {Dietrich},
  {Dodelson}, {Doel}, {Drlica-Wagner}, {Eifler}, {Elliott}, {Elsner},
  {Elvin-Poole}, {Estrada}, {Evrard}, {Fang}, {Fernandez}, {Fert{\'e}},
  {Finley}, {Flaugher}, {Fosalba}, {Friedrich}, {Frieman},
  {Garc{\'\i}a-Bellido}, {Garcia-Fernandez}, {Gatti}, {Gaztanaga}, {Gerdes},
  {Giannantonio}, {Gill}, {Glazebrook}, {Goldstein}, {Gruen}, {Gruendl},
  {Gschwend}, {Gutierrez}, {Hamilton}, {Hartley}, {Hinton}, {Honscheid},
  {Hoyle}, {Huterer}, {Jain}, {James}, {Jarvis}, {Jeltema}, {Johnson},
  {Johnson}, {Kacprzak}, {Kent}, {Kim}, {King}, {Kirk}, {Kokron}, {Kovacs},
  {Krause}, {Krawiec}, {Kremin}, {Kuehn}, {Kuhlmann}, {Kuropatkin}, {Lacasa},
  {Lahav}, {Li}, {Liddle}, {Lidman}, {Lima}, {Lin}, {MacCrann}, {Maia},
  {Makler}, {Manera}, {March}, {Marshall}, {Martini}, {McMahon}, {Melchior},
  {Menanteau}, {Miquel}, {Miranda}, {Mudd}, {Muir}, {M{\"o}ller}, {Neilsen},
  {Nichol}, {Nord}, {Nugent}, {Ogando}, {Palmese}, {Peacock}, {Peiris},
  {Peoples}, {Percival}, {Petravick}, {Plazas}, {Porredon}, {Prat}, {Pujol},
  {Rau}, {Refregier}, {Ricker}, {Roe}, {Rollins}, {Romer}, {Roodman},
  {Rosenfeld}, {Ross}, {Rozo}, {Rykoff}, {Sako}, {Salvador}, {Samuroff},
  {S{\'a}nchez}, {Sanchez}, {Santiago}, {Scarpine}, {Schindler}, {Scolnic},
  {Secco}, {Serrano}, {Sevilla-Noarbe}, {Sheldon}, {Smith}, {Smith}, {Smith},
  {Soares-Santos}, {Sobreira}, {Suchyta}, {Tarle}, {Thomas}, {Troxel},
  {Tucker}, {Tucker}, {Uddin}, {Varga}, {Vielzeuf}, {Vikram}, {Vivas},
  {Walker}, {Wang}, {Wechsler}, {Weller}, {Wester}, {Wolf}, {Yanny}, {Yuan},
  {Zenteno}, {Zhang}, {Zhang}, {Zuntz}, \& {Dark Energy Survey
  Collaboration}}]{2018PhRvD..98d3526A}
---. 2018{\natexlab{b}}, \prd, 98, 043526, \dodoi{10.1103/PhysRevD.98.043526}

\bibitem[{{Abbott} {et~al.}(2022){Abbott}, {Aguena}, {Alarcon}, {Allam},
  {Alves}, {Amon}, {Andrade-Oliveira}, {Annis}, {Avila}, {Bacon}, {Baxter},
  {Bechtol}, {Becker}, {Bernstein}, {Bhargava}, {Birrer}, {Blazek},
  {Brandao-Souza}, {Bridle}, {Brooks}, {Buckley-Geer}, {Burke}, {Camacho},
  {Campos}, {Carnero Rosell}, {Carrasco Kind}, {Carretero}, {Castander},
  {Cawthon}, {Chang}, {Chen}, {Chen}, {Choi}, {Conselice}, {Cordero},
  {Costanzi}, {Crocce}, {da Costa}, {da Silva Pereira}, {Davis}, {Davis}, {De
  Vicente}, {DeRose}, {Desai}, {Di Valentino}, {Diehl}, {Dietrich}, {Dodelson},
  {Doel}, {Doux}, {Drlica-Wagner}, {Eckert}, {Eifler}, {Elsner}, {Elvin-Poole},
  {Everett}, {Evrard}, {Fang}, {Farahi}, {Fernandez}, {Ferrero}, {Fert{\'e}},
  {Fosalba}, {Friedrich}, {Frieman}, {Garc{\'\i}a-Bellido}, {Gatti},
  {Gaztanaga}, {Gerdes}, {Giannantonio}, {Giannini}, {Gruen}, {Gruendl},
  {Gschwend}, {Gutierrez}, {Harrison}, {Hartley}, {Herner}, {Hinton},
  {Hollowood}, {Honscheid}, {Hoyle}, {Huff}, {Huterer}, {Jain}, {James},
  {Jarvis}, {Jeffrey}, {Jeltema}, {Kovacs}, {Krause}, {Kron}, {Kuehn},
  {Kuropatkin}, {Lahav}, {Leget}, {Lemos}, {Liddle}, {Lidman}, {Lima}, {Lin},
  {MacCrann}, {Maia}, {Marshall}, {Martini}, {McCullough}, {Melchior},
  {Mena-Fern{\'a}ndez}, {Menanteau}, {Miquel}, {Mohr}, {Morgan}, {Muir},
  {Myles}, {Nadathur}, {Navarro-Alsina}, {Nichol}, {Ogando}, {Omori},
  {Palmese}, {Pandey}, {Park}, {Paz-Chinch{\'o}n}, {Petravick}, {Pieres},
  {Plazas Malag{\'o}n}, {Porredon}, {Prat}, {Raveri}, {Rodriguez-Monroy},
  {Rollins}, {Romer}, {Roodman}, {Rosenfeld}, {Ross}, {Rykoff}, {Samuroff},
  {S{\'a}nchez}, {Sanchez}, {Sanchez}, {Sanchez Cid}, {Scarpine}, {Schubnell},
  {Scolnic}, {Secco}, {Serrano}, {Sevilla-Noarbe}, {Sheldon}, {Shin}, {Smith},
  {Soares-Santos}, {Suchyta}, {Swanson}, {Tabbutt}, {Tarle}, {Thomas}, {To},
  {Troja}, {Troxel}, {Tucker}, {Tutusaus}, {Varga}, {Walker}, {Weaverdyck},
  {Wechsler}, {Weller}, {Yanny}, {Yin}, {Zhang}, {Zuntz}, \& {DES
  Collaboration}}]{DESreal2}
{Abbott}, T.~M.~C., {Aguena}, M., {Alarcon}, A., {et~al.} 2022, \prd, 105,
  023520, \dodoi{10.1103/PhysRevD.105.023520}

\bibitem[{{Akeson} {et~al.}(2019){Akeson}, {Armus}, {Bachelet}, {Bailey},
  {Bartusek}, {Bellini}, {Benford}, {Bennett}, {Bhattacharya}, {Bohlin},
  {Boyer}, {Bozza}, {Bryden}, {Calchi Novati}, {Carpenter}, {Casertano},
  {Choi}, {Content}, {Dayal}, {Dressler}, {Dor{\'e}}, {Fall}, {Fan}, {Fang},
  {Filippenko}, {Finkelstein}, {Foley}, {Furlanetto}, {Kalirai}, {Gaudi},
  {Gilbert}, {Girard}, {Grady}, {Greene}, {Guhathakurta}, {Heinrich},
  {Hemmati}, {Hendel}, {Henderson}, {Henning}, {Hirata}, {Ho}, {Huff},
  {Hutter}, {Jansen}, {Jha}, {Johnson}, {Jones}, {Kasdin}, {Kelly}, {Kirshner},
  {Koekemoer}, {Kruk}, {Lewis}, {Macintosh}, {Madau}, {Malhotra}, {Mandel},
  {Massara}, {Masters}, {McEnery}, {McQuinn}, {Melchior}, {Melton},
  {Mennesson}, {Peeples}, {Penny}, {Perlmutter}, {Pisani}, {Plazas}, {Poleski},
  {Postman}, {Ranc}, {Rauscher}, {Rest}, {Roberge}, {Robertson}, {Rodney},
  {Rhoads}, {Rhodes}, {Ryan}, {Sahu}, {Sand}, {Scolnic}, {Seth}, {Shvartzvald},
  {Siellez}, {Smith}, {Spergel}, {Stassun}, {Street}, {Strolger}, {Szalay},
  {Trauger}, {Troxel}, {Turnbull}, {van der Marel}, {von der Linden}, {Wang},
  {Weinberg}, {Williams}, {Windhorst}, {Wollack}, {Wu}, {Yee}, \&
  {Zimmerman}}]{WFIRST}
{Akeson}, R., {Armus}, L., {Bachelet}, E., {et~al.} 2019, arXiv e-prints,
  arXiv:1902.05569, \dodoi{10.48550/arXiv.1902.05569}

\bibitem[{{Alam} {et~al.}(2017){Alam}, {Ata}, {Bailey}, {Beutler}, {Bizyaev},
  {Blazek}, {Bolton}, {Brownstein}, {Burden}, {Chuang}, {Comparat}, {Cuesta},
  {Dawson}, {Eisenstein}, {Escoffier}, {Gil-Mar{\'\i}n}, {Grieb}, {Hand}, {Ho},
  {Kinemuchi}, {Kirkby}, {Kitaura}, {Malanushenko}, {Malanushenko}, {Maraston},
  {McBride}, {Nichol}, {Olmstead}, {Oravetz}, {Padmanabhan},
  {Palanque-Delabrouille}, {Pan}, {Pellejero-Ibanez}, {Percival}, {Petitjean},
  {Prada}, {Price-Whelan}, {Reid}, {Rodr{\'\i}guez-Torres}, {Roe}, {Ross},
  {Ross}, {Rossi}, {Rubi{\~n}o-Mart{\'\i}n}, {Saito}, {Salazar-Albornoz},
  {Samushia}, {S{\'a}nchez}, {Satpathy}, {Schlegel}, {Schneider},
  {Sc{\'o}ccola}, {Seo}, {Sheldon}, {Simmons}, {Slosar}, {Strauss}, {Swanson},
  {Thomas}, {Tinker}, {Tojeiro}, {Maga{\~n}a}, {Vazquez}, {Verde}, {Wake},
  {Wang}, {Weinberg}, {White}, {Wood-Vasey}, {Y{\`e}che}, {Zehavi}, {Zhai}, \&
  {Zhao}}]{BAO1}
{Alam}, S., {Ata}, M., {Bailey}, S., {et~al.} 2017, \mnras, 470, 2617,
  \dodoi{10.1093/mnras/stx721}

\bibitem[{{Alonso} {et~al.}(2019){Alonso}, {Sanchez}, {Slosar}, \& {LSST Dark
  Energy Science Collaboration}}]{2019MNRAS.484.4127A}
{Alonso}, D., {Sanchez}, J., {Slosar}, A., \& {LSST Dark Energy Science
  Collaboration}. 2019, \mnras, 484, 4127, \dodoi{10.1093/mnras/stz093}

\bibitem[{{Asgari} {et~al.}(2021){Asgari}, {Lin}, {Joachimi}, {Giblin},
  {Heymans}, {Hildebrandt}, {Kannawadi}, {St{\"o}lzner}, {Tr{\"o}ster}, {van
  den Busch}, {Wright}, {Bilicki}, {Blake}, {de Jong}, {Dvornik}, {Erben},
  {Getman}, {Hoekstra}, {K{\"o}hlinger}, {Kuijken}, {Miller}, {Radovich},
  {Schneider}, {Shan}, \& {Valentijn}}]{S81}
{Asgari}, M., {Lin}, C.-A., {Joachimi}, B., {et~al.} 2021, \aap, 645, A104,
  \dodoi{10.1051/0004-6361/202039070}

\bibitem[{{Bernstein}(2009)}]{3x2pt1}
{Bernstein}, G.~M. 2009, \apj, 695, 652, \dodoi{10.1088/0004-637X/695/1/652}

\bibitem[{{Blaizot} {et~al.}(2005){Blaizot}, {Wadadekar}, {Guiderdoni},
  {Colombi}, {Bertin}, {Bouchet}, {Devriendt}, \&
  {Hatton}}]{2005MNRAS.360..159B}
{Blaizot}, J., {Wadadekar}, Y., {Guiderdoni}, B., {et~al.} 2005, \mnras, 360,
  159, \dodoi{10.1111/j.1365-2966.2005.09019.x}

\bibitem[{{Bruzual} \& {Charlot}(2003)}]{BC03}
{Bruzual}, G., \& {Charlot}, S. 2003, \mnras, 344, 1000,
  \dodoi{10.1046/j.1365-8711.2003.06897.x}

\bibitem[{{Cao} {et~al.}(2018){Cao}, {Gong}, {Meng}, {Xu}, {Chen}, {Guo}, {Li},
  {Liu}, {Xue}, {Cao}, {Fu}, {Zhang}, {Wang}, \& {Zhan}}]{cao}
{Cao}, Y., {Gong}, Y., {Meng}, X.-M., {et~al.} 2018, \mnras, 480, 2178,
  \dodoi{10.1093/mnras/sty1980}

\bibitem[{{Capak} {et~al.}(2007){Capak}, {Aussel}, {Ajiki}, {McCracken},
  {Mobasher}, {Scoville}, {Shopbell}, {Taniguchi}, {Thompson}, {Tribiano},
  {Sasaki}, {Blain}, {Brusa}, {Carilli}, {Comastri}, {Carollo}, {Cassata},
  {Colbert}, {Ellis}, {Elvis}, {Giavalisco}, {Green}, {Guzzo}, {Hasinger},
  {Ilbert}, {Impey}, {Jahnke}, {Kartaltepe}, {Kneib}, {Koda}, {Koekemoer},
  {Komiyama}, {Leauthaud}, {Le Fevre}, {Lilly}, {Liu}, {Massey}, {Miyazaki},
  {Murayama}, {Nagao}, {Peacock}, {Pickles}, {Porciani}, {Renzini}, {Rhodes},
  {Rich}, {Salvato}, {Sanders}, {Scarlata}, {Schiminovich}, {Schinnerer},
  {Scodeggio}, {Sheth}, {Shioya}, {Tasca}, {Taylor}, {Yan}, \&
  {Zamorani}}]{2007ApJS..172...99C}
{Capak}, P., {Aussel}, H., {Ajiki}, M., {et~al.} 2007, \apjs, 172, 99,
  \dodoi{10.1086/519081}

\bibitem[{{Carlson} \& {White}(2010)}]{2010ApJS..190..311C}
{Carlson}, J., \& {White}, M. 2010, \apjs, 190, 311,
  \dodoi{10.1088/0067-0049/190/2/311}

\bibitem[{{Cawthon} {et~al.}(2022){Cawthon}, {Elvin-Poole}, {Porredon},
  {Crocce}, {Giannini}, {Gatti}, {Ross}, {Rykoff}, {Carnero Rosell}, {DeRose},
  {Lee}, {Rodriguez-Monroy}, {Amon}, {Bechtol}, {De Vicente}, {Gruen},
  {Morgan}, {Sanchez}, {Sanchez}, {Sevilla-Noarbe}, {Abbott}, {Aguena},
  {Allam}, {Annis}, {Avila}, {Bacon}, {Bertin}, {Brooks}, {Burke}, {Carrasco
  Kind}, {Carretero}, {Castander}, {Choi}, {Costanzi}, {da Costa}, {Pereira},
  {Dawson}, {Desai}, {Diehl}, {Eckert}, {Everett}, {Ferrero}, {Fosalba},
  {Frieman}, {Garc{\'\i}a-Bellido}, {Gaztanaga}, {Gruendl}, {Gschwend},
  {Gutierrez}, {Hinton}, {Hollowood}, {Honscheid}, {Huterer}, {James}, {Kim},
  {Kneib}, {Kuehn}, {Kuropatkin}, {Lahav}, {Lima}, {Lin}, {Maia}, {Melchior},
  {Menanteau}, {Miquel}, {Mohr}, {Muir}, {Myles}, {Palmese}, {Pandey},
  {Paz-Chinch{\'o}n}, {Percival}, {Plazas}, {Roodman}, {Rossi}, {Scarpine},
  {Serrano}, {Smith}, {Soares-Santos}, {Suchyta}, {Swanson}, {Tarle}, {To},
  {Troxel}, {Wilkinson}, \& {DES Collaboration}}]{redun1}
{Cawthon}, R., {Elvin-Poole}, J., {Porredon}, A., {et~al.} 2022, \mnras, 513,
  5517, \dodoi{10.1093/mnras/stac1160}

\bibitem[{{Chabrier}(2003)}]{Chabrier}
{Chabrier}, G. 2003, \pasp, 115, 763, \dodoi{10.1086/376392}

\bibitem[{{Chang} {et~al.}(2016){Chang}, {Pujol}, {Gazta{\~n}aga}, {Amara},
  {R{\'e}fr{\'e}gier}, {Bacon}, {Becker}, {Bonnett}, {Carretero}, {Castander},
  {Crocce}, {Fosalba}, {Giannantonio}, {Hartley}, {Jarvis}, {Kacprzak}, {Ross},
  {Sheldon}, {Troxel}, {Vikram}, {Zuntz}, {Abbott}, {Abdalla}, {Allam},
  {Annis}, {Benoit-L{\'e}vy}, {Bertin}, {Brooks}, {Buckley-Geer}, {Burke},
  {Capozzi}, {Carnero Rosell}, {Carrasco Kind}, {Cunha}, {D'Andrea}, {da
  Costa}, {Desai}, {Diehl}, {Dietrich}, {Doel}, {Eifler}, {Estrada}, {Evrard},
  {Flaugher}, {Frieman}, {Goldstein}, {Gruen}, {Gruendl}, {Gutierrez},
  {Honscheid}, {Jain}, {James}, {Kuehn}, {Kuropatkin}, {Lahav}, {Li}, {Lima},
  {Marshall}, {Martini}, {Melchior}, {Miller}, {Miquel}, {Mohr}, {Nichol},
  {Nord}, {Ogando}, {Plazas}, {Reil}, {Romer}, {Roodman}, {Rykoff}, {Sanchez},
  {Scarpine}, {Schubnell}, {Sevilla-Noarbe}, {Smith}, {Soares-Santos},
  {Sobreira}, {Suchyta}, {Swanson}, {Tarle}, {Thomas}, \& {Walker}}]{gbias}
{Chang}, C., {Pujol}, A., {Gazta{\~n}aga}, E., {et~al.} 2016, \mnras, 459,
  3203, \dodoi{10.1093/mnras/stw861}

\bibitem[{{Charlot} \& {Fall}(2000)}]{Charlot2000ApJ...539..718C}
{Charlot}, S., \& {Fall}, S.~M. 2000, \apj, 539, 718, \dodoi{10.1086/309250}

\bibitem[{{Chen} \& {Yu}(2024)}]{cz}
{Chen}, Z., \& {Yu}, Y. 2024, \mnras, 534, 1205, \dodoi{10.1093/mnras/stae2150}

\bibitem[{{Chisari} {et~al.}(2019){Chisari}, {Alonso}, {Krause}, {Leonard},
  {Bull}, {Neveu}, {Villarreal}, {Singh}, {McClintock}, {Ellison}, {Du},
  {Zuntz}, {Mead}, {Joudaki}, {Lorenz}, {Tr{\"o}ster}, {Sanchez}, {Lanusse},
  {Ishak}, {Hlozek}, {Blazek}, {Campagne}, {Almoubayyed}, {Eifler}, {Kirby},
  {Kirkby}, {Plaszczynski}, {Slosar}, {Vrastil}, {Wagoner}, \& {LSST Dark
  Energy Science Collaboration}}]{CCL}
{Chisari}, N.~E., {Alonso}, D., {Krause}, E., {et~al.} 2019, \apjs, 242, 2,
  \dodoi{10.3847/1538-4365/ab1658}

\bibitem[{{Chon} {et~al.}(2004){Chon}, {Challinor}, {Prunet}, {Hivon}, \&
  {Szapudi}}]{pseudo3}
{Chon}, G., {Challinor}, A., {Prunet}, S., {Hivon}, E., \& {Szapudi}, I. 2004,
  \mnras, 350, 914, \dodoi{10.1111/j.1365-2966.2004.07737.x}

\bibitem[{{Dark Energy Survey and Kilo-Degree Survey Collaboration}
  {et~al.}(2023){Dark Energy Survey and Kilo-Degree Survey Collaboration},
  {Abbott}, {Aguena}, {Alarcon}, {Alves}, {Amon}, {Andrade-Oliveira}, {Asgari},
  {Avila}, {Bacon}, {Bechtol}, {Becker}, {Bernstein}, {Bertin}, {Bilicki},
  {Blazek}, {Bocquet}, {Brooks}, {Burger}, {Burke}, {Camacho}, {Campos},
  {Carnero Rosell}, {Carrasco Kind}, {Carretero}, {Castander}, {Cawthon},
  {Chang}, {Chen}, {Choi}, {Conselice}, {Cordero}, {Crocce}, {da Costa}, {da
  Silva Pereira}, {Dalal}, {Davis}, {de Jong}, {DeRose}, {Desai}, {Diehl},
  {Dodelson}, {Doel}, {Doux}, {Drlica-Wagner}, {Dvornik}, {Eckert}, {Eifler},
  {Elvin-Poole}, {Everett}, {Fang}, {Ferrero}, {Fert{\'e}}, {Flaugher},
  {Friedrich}, {Frieman}, {Garc{\'\i}a-Bellido}, {Gatti}, {Giannini}, {Giblin},
  {Gruen}, {Gruendl}, {Gutierrez}, {Harrison}, {Hartley}, {Herner}, {Heymans},
  {Hildebrandt}, {Hinton}, {Hoekstra}, {Hollowood}, {Honscheid}, {Huang},
  {Huff}, {Huterer}, {James}, {Jarvis}, {Jeffrey}, {Jeltema}, {Joachimi},
  {Joudaki}, {Kannawadi}, {Krause}, {Kuehn}, {Kuijken}, {Kuropatkin}, {Lahav},
  {Leget}, {Lemos}, {Li}, {Li}, {Liddle}, {Lima}, {Lin}, {Lin}, {MacCrann},
  {Mahony}, {Marshall}, {McCullough}, {Mena-Fern{\'a}ndez}, {Menanteau},
  {Miquel}, {Mohr}, {Muir}, {Myles}, {Napolitano}, {Navarro-Alsina}, {Ogando},
  {Palmese}, {Pandey}, {Park}, {Paterno}, {Peacock}, {Petravick}, {Pieres},
  {Plazas Malag{\'o}n}, {Porredon}, {Prat}, {Radovich}, {Raveri}, {Reischke},
  {Robertson}, {Rollins}, {Romer}, {Roodman}, {Rykoff}, {Samuroff},
  {S{\'a}nchez}, {Sanchez}, {Sanchez}, {Schneider}, {Secco}, {Sevilla-Noarbe},
  {Shan}, {Sheldon}, {Shin}, {Sif{\'o}n}, {Smith}, {Soares-Santos},
  {St{\"o}lzner}, {Suchyta}, {Swanson}, {Tarle}, {Thomas}, {To}, {Troxel},
  {Tr{\"o}ster}, {Tutusaus}, {van den Busch}, {Varga}, {Walker}, {Weaverdyck},
  {Wechsler}, {Weller}, {Wiseman}, {Wright}, {Yanny}, {Yin}, {Yoon}, {Zhang},
  \& {Zuntz}}]{2023OJAp....6E..36D}
{Dark Energy Survey and Kilo-Degree Survey Collaboration}, {Abbott}, T.~M.~C.,
  {Aguena}, M., {et~al.} 2023, The Open Journal of Astrophysics, 6, 36,
  \dodoi{10.21105/astro.2305.17173}

\bibitem[{{Devriendt} {et~al.}(1999){Devriendt}, {Guiderdoni}, \&
  {Sadat}}]{Devriendt}
{Devriendt}, J.~E.~G., {Guiderdoni}, B., \& {Sadat}, R. 1999, \aap, 350, 381,
  \dodoi{10.48550/arXiv.astro-ph/9906332}

\bibitem[{{Duffy} {et~al.}(2008){Duffy}, {Schaye}, {Kay}, \& {Dalla
  Vecchia}}]{c-m}
{Duffy}, A.~R., {Schaye}, J., {Kay}, S.~T., \& {Dalla Vecchia}, C. 2008,
  \mnras, 390, L64, \dodoi{10.1111/j.1745-3933.2008.00537.x}

\bibitem[{{Eifler} {et~al.}(2021){Eifler}, {Simet}, {Krause}, {Hirata},
  {Huang}, {Fang}, {Miranda}, {Mandelbaum}, {Doux}, {Heinrich}, {Huff},
  {Miyatake}, {Hemmati}, {Xu}, {Rogozenski}, {Capak}, {Choi}, {Dor{\'e}},
  {Jain}, {Jarvis}, {Kruk}, {MacCrann}, {Masters}, {Rozo}, {Spergel}, {Troxel},
  {von der Linden}, {Wang}, {Weinberg}, {Wenzl}, \& {Wu}}]{3x2pt5}
{Eifler}, T., {Simet}, M., {Krause}, E., {et~al.} 2021, \mnras, 507, 1514,
  \dodoi{10.1093/mnras/stab533}

\bibitem[{Elad {et~al.}(2005)Elad, Starck, Querre, \& Donoho}]{inpainting1}
Elad, M., Starck, J.-L., Querre, P., \& Donoho, D.~L. 2005, Applied and
  computational harmonic analysis, 19, 340

\bibitem[{{Euclid Collaboration} {et~al.}(2020){Euclid Collaboration},
  {Blanchard}, {Camera}, {Carbone}, {Cardone}, {Casas}, {Clesse}, {Ili{\'c}},
  {Kilbinger}, {Kitching}, {Kunz}, {Lacasa}, {Linder}, {Majerotto},
  {Markovi{\v{c}}}, {Martinelli}, {Pettorino}, {Pourtsidou}, {Sakr},
  {S{\'a}nchez}, {Sapone}, {Tutusaus}, {Yahia-Cherif}, {Yankelevich},
  {Andreon}, {Aussel}, {Balaguera-Antol{\'\i}nez}, {Baldi}, {Bardelli},
  {Bender}, {Biviano}, {Bonino}, {Boucaud}, {Bozzo}, {Branchini}, {Brau-Nogue},
  {Brescia}, {Brinchmann}, {Burigana}, {Cabanac}, {Capobianco}, {Cappi},
  {Carretero}, {Carvalho}, {Casas}, {Castander}, {Castellano}, {Cavuoti},
  {Cimatti}, {Cledassou}, {Colodro-Conde}, {Congedo}, {Conselice}, {Conversi},
  {Copin}, {Corcione}, {Coupon}, {Courtois}, {Cropper}, {Da Silva}, {de la
  Torre}, {Di Ferdinando}, {Dubath}, {Ducret}, {Duncan}, {Dupac}, {Dusini},
  {Fabbian}, {Fabricius}, {Farrens}, {Fosalba}, {Fotopoulou}, {Fourmanoit},
  {Frailis}, {Franceschi}, {Franzetti}, {Fumana}, {Galeotta}, {Gillard},
  {Gillis}, {Giocoli}, {G{\'o}mez-Alvarez}, {Graci{\'a}-Carpio}, {Grupp},
  {Guzzo}, {Hoekstra}, {Hormuth}, {Israel}, {Jahnke}, {Keihanen}, {Kermiche},
  {Kirkpatrick}, {Kohley}, {Kubik}, {Kurki-Suonio}, {Ligori}, {Lilje}, {Lloro},
  {Maino}, {Maiorano}, {Marggraf}, {Martinet}, {Marulli}, {Massey},
  {Medinaceli}, {Mei}, {Mellier}, {Metcalf}, {Metge}, {Meylan}, {Moresco},
  {Moscardini}, {Munari}, {Nichol}, {Niemi}, {Nucita}, {Padilla}, {Paltani},
  {Pasian}, {Percival}, {Pires}, {Polenta}, {Poncet}, {Pozzetti}, {Racca},
  {Raison}, {Renzi}, {Rhodes}, {Romelli}, {Roncarelli}, {Rossetti}, {Saglia},
  {Schneider}, {Scottez}, {Secroun}, {Sirri}, {Stanco}, {Starck}, {Sureau},
  {Tallada-Cresp{\'\i}}, {Tavagnacco}, {Taylor}, {Tenti}, {Tereno},
  {Toledo-Moreo}, {Torradeflot}, {Valenziano}, {Vassallo}, {Verdoes Kleijn},
  {Viel}, {Wang}, {Zacchei}, {Zoubian}, \& {Zucca}}]{3x2pt6}
{Euclid Collaboration}, {Blanchard}, A., {Camera}, S., {et~al.} 2020, \aap,
  642, A191, \dodoi{10.1051/0004-6361/202038071}

\bibitem[{{Euclid Collaboration} {et~al.}(2024){Euclid Collaboration},
  {Mellier}, {Abdurro'uf}, {Acevedo Barroso}, {Ach{\'u}carro}, {Adamek},
  {Adam}, {Addison}, {Aghanim}, {Aguena}, {Ajani}, {Akrami}, {Al-Bahlawan},
  {Alavi}, {Albuquerque}, {Alestas}, {Alguero}, {Allaoui}, {Allen}, {Allevato},
  {Alonso-Tetilla}, {Altieri}, {Alvarez-Candal}, {Alvi}, {Amara}, {Amendola},
  {Amiaux}, {Andika}, {Andreon}, {Andrews}, {Angora}, {Angulo}, {Annibali},
  {Anselmi}, {Anselmi}, {Arcari}, {Archidiacono}, {Aric{\`o}}, {Arnaud},
  {Arnouts}, {Asgari}, {Asorey}, {Atayde}, {Atek}, {Atrio-Barandela}, {Aubert},
  {Aubourg}, {Auphan}, {Auricchio}, {Aussel}, {Aussel}, {Avelino},
  {Avgoustidis}, {Avila}, {Awan}, {Azzollini}, {Baccigalupi}, {Bachelet},
  {Bacon}, {Baes}, {Bagley}, {Bahr-Kalus}, {Balaguera-Antolinez}, {Balbinot},
  {Balcells}, {Baldi}, {Baldry}, {Balestra}, {Ballardini}, {Ballester},
  {Balogh}, {Ba{\~n}ados}, {Barbier}, {Bardelli}, {Baron}, {Barreiro},
  {Barrena}, {Barriere}, {Barros}, {Barthelemy}, {Bartolo}, {Basset},
  {Battaglia}, {Battisti}, {Baugh}, {Baumont}, {Bazzanini}, {Beaulieu},
  {Beckmann}, {Belikov}, {Bel}, {Bellagamba}, {Bella}, {Bellini}, {Benabed},
  {Bender}, {Benevento}, {Bennett}, {Benson}, {Bergamini}, {Bermejo-Climent},
  {Bernardeau}, {Bertacca}, {Berthe}, {Berthier}, {Bethermin}, {Beutler},
  {Bevillon}, {Bhargava}, {Bhatawdekar}, {Bianchi}, {Bisigello}, {Biviano},
  {Blake}, {Blanchard}, {Blazek}, {Blot}, {Bosco}, {Bodendorf}, {Boenke},
  {B{\"o}hringer}, {Boldrini}, {Bolzonella}, {Bonchi}, {Bonici}, {Bonino},
  {Bonino}, {Bonvin}, {Bon}, {Booth}, {Borgani}, {Borlaff}, {Borsato}, {Bosco},
  {Bose}, {Botticella}, {Boucaud}, {Bouche}, {Boucher}, {Boutigny}, {Bouvard},
  {Bouwens}, {Bouy}, {Bowler}, {Bozza}, {Bozzo}, {Branchini}, {Brando},
  {Brau-Nogue}, {Brekke}, {Bremer}, {Brescia}, {Breton}, {Brinchmann},
  {Brinckmann}, {Brockley-Blatt}, {Brodwin}, {Brouard}, {Brown}, {Bruton},
  {Bucko}, {Buddelmeijer}, {Buenadicha}, {Buitrago}, {Burger}, {Burigana},
  {Busillo}, {Busonero}, {Cabanac}, {Cabayol-Garcia}, {Cagliari}, {Caillat},
  {Caillat}, {Calabrese}, {Calabro}, {Calderone}, {Calura}, {Camacho Quevedo},
  {Camera}, {Campos}, {Canas-Herrera}, {Candini}, {Cantiello}, {Capobianco},
  {Cappellaro}, {Cappelluti}, {Cappi}, {Caputi}, {Cara}, {Carbone}, {Cardone},
  {Carella}, {Carlberg}, {Carle}, {Carminati}, {Caro}, {Carrasco}, {Carretero},
  {Carrilho}, \& {Carron Duque}}]{Euclid}
{Euclid Collaboration}, {Mellier}, Y., {Abdurro'uf}, {et~al.} 2024, arXiv
  e-prints, arXiv:2405.13491, \dodoi{10.48550/arXiv.2405.13491}

\bibitem[{{Faga} {et~al.}(2024){Faga}, {Andrade-Oliveira}, {Camacho},
  {Rosenfeld}, {Lima}, {Doux}, {Fang}, {Prat}, {Porredon}, {Aguena}, {Alarcon},
  {Allam}, {Alves}, {Amon}, {Avila}, {Bacon}, {Bechtol}, {Becker}, {Bernstein},
  {Bocquet}, {Brooks}, {Buckley-Geer}, {Campos}, {Carnero Rosell}, {Carrasco
  Kind}, {Carretero}, {Castander}, {Cawthon}, {Chang}, {Chen}, {Choi},
  {Cordero}, {Crocce}, {da Costa}, {Pereira}, {DeRose}, {Diehl}, {Dodelson},
  {Drlica-Wagner}, {Elvin-Poole}, {Everett}, {Ferrero}, {Fert{\'e}},
  {Flaugher}, {Fosalba}, {Frieman}, {Garc{\'\i}a-Bellido}, {Gatti},
  {Gaztanaga}, {Giannini}, {Gruen}, {Gruendl}, {Gutierrez}, {Harrison},
  {Hinton}, {Hollowood}, {Honscheid}, {Huterer}, {James}, {Jarvis}, {Jeltema},
  {Kuehn}, {Lahav}, {Lee}, {Lidman}, {MacCrann}, {Marshall}, {McCullough},
  {Mena-Fern{\'a}ndez}, {Miquel}, {Myles}, {Navarro-Alsina}, {Palmese},
  {Pandey}, {Paterno}, {Pieres}, {Plazas Malag{\'o}n}, {Raveri},
  {Rodriguez-Monroy}, {Rollins}, {Ross}, {Rykoff}, {Samuroff}, {S{\'a}nchez},
  {Sanchez}, {Sanchez Cid}, {Schubnell}, {Secco}, {Sevilla-Noarbe}, {Sheldon},
  {Shin}, {Smith}, {Soares-Santos}, {Suchyta}, {Swanson}, {Tarle}, {Thomas},
  {Troxel}, {Tutusaus}, {Weaverdyck}, {Wiseman}, {Yanny}, \& {Yin}}]{comres1}
{Faga}, L., {Andrade-Oliveira}, F., {Camacho}, H., {et~al.} 2024, arXiv
  e-prints, arXiv:2406.12675, \dodoi{10.48550/arXiv.2406.12675}

\bibitem[{{Foreman-Mackey} {et~al.}(2013{\natexlab{a}}){Foreman-Mackey},
  {Hogg}, {Lang}, \& {Goodman}}]{emcee}
{Foreman-Mackey}, D., {Hogg}, D.~W., {Lang}, D., \& {Goodman}, J.
  2013{\natexlab{a}}, PASP, 125, 306, \dodoi{10.1086/670067}

\bibitem[{{Foreman-Mackey} {et~al.}(2013{\natexlab{b}}){Foreman-Mackey},
  {Hogg}, {Lang}, \& {Goodman}}]{2013PASP..125..306F}
---. 2013{\natexlab{b}}, \pasp, 125, 306, \dodoi{10.1086/670067}

\bibitem[{{Frieman} {et~al.}(2008){Frieman}, {Turner}, \& {Huterer}}]{LCDM1}
{Frieman}, J.~A., {Turner}, M.~S., \& {Huterer}, D. 2008, \araa, 46, 385,
  \dodoi{10.1146/annurev.astro.46.060407.145243}

\bibitem[{{Gong} {et~al.}(2019){Gong}, {Liu}, {Cao}, {Chen}, {Fan}, {Li}, {Li},
  {Li}, {Zhang}, \& {Zhan}}]{gong}
{Gong}, Y., {Liu}, X., {Cao}, Y., {et~al.} 2019, \apj, 883, 203,
  \dodoi{10.3847/1538-4357/ab391e}

\bibitem[{{Hamana} {et~al.}(2004){Hamana}, {Takada}, \& {Yoshida}}]{randsig}
{Hamana}, T., {Takada}, M., \& {Yoshida}, N. 2004, \mnras, 350, 893,
  \dodoi{10.1111/j.1365-2966.2004.07691.x}

\bibitem[{{Hamana} {et~al.}(2020){Hamana}, {Shirasaki}, {Miyazaki}, {Hikage},
  {Oguri}, {More}, {Armstrong}, {Leauthaud}, {Mandelbaum}, {Miyatake},
  {Nishizawa}, {Simet}, {Takada}, {Aihara}, {Bosch}, {Komiyama}, {Lupton},
  {Murayama}, {Strauss}, \& {Tanaka}}]{2020PASJ...72...16H}
{Hamana}, T., {Shirasaki}, M., {Miyazaki}, S., {et~al.} 2020, \pasj, 72, 16,
  \dodoi{10.1093/pasj/psz138}

\bibitem[{{Henriques} {et~al.}(2015){Henriques}, {White}, {Thomas}, {Angulo},
  {Guo}, {Lemson}, {Springel}, \& {Overzier}}]{henriques2015galaxy}
{Henriques}, B. M.~B., {White}, S. D.~M., {Thomas}, P.~A., {et~al.} 2015,
  \mnras, 451, 2663, \dodoi{10.1093/mnras/stv705}

\bibitem[{{Heymans} {et~al.}(2021{\natexlab{a}}){Heymans}, {Tr{\"o}ster},
  {Asgari}, {Blake}, {Hildebrandt}, {Joachimi}, {Kuijken}, {Lin},
  {S{\'a}nchez}, {van den Busch}, {Wright}, {Amon}, {Bilicki}, {de Jong},
  {Crocce}, {Dvornik}, {Erben}, {Fortuna}, {Getman}, {Giblin}, {Glazebrook},
  {Hoekstra}, {Joudaki}, {Kannawadi}, {K{\"o}hlinger}, {Lidman}, {Miller},
  {Napolitano}, {Parkinson}, {Schneider}, {Shan}, {Valentijn}, {Verdoes
  Kleijn}, \& {Wolf}}]{KiDsfour2}
{Heymans}, C., {Tr{\"o}ster}, T., {Asgari}, M., {et~al.} 2021{\natexlab{a}},
  \aap, 646, A140, \dodoi{10.1051/0004-6361/202039063}

\bibitem[{{Heymans} {et~al.}(2021{\natexlab{b}}){Heymans}, {Tr{\"o}ster},
  {Asgari}, {Blake}, {Hildebrandt}, {Joachimi}, {Kuijken}, {Lin},
  {S{\'a}nchez}, {van den Busch}, {Wright}, {Amon}, {Bilicki}, {de Jong},
  {Crocce}, {Dvornik}, {Erben}, {Fortuna}, {Getman}, {Giblin}, {Glazebrook},
  {Hoekstra}, {Joudaki}, {Kannawadi}, {K{\"o}hlinger}, {Lidman}, {Miller},
  {Napolitano}, {Parkinson}, {Schneider}, {Shan}, {Valentijn}, {Verdoes
  Kleijn}, \& {Wolf}}]{2021A&A...646A.140H}
---. 2021{\natexlab{b}}, \aap, 646, A140, \dodoi{10.1051/0004-6361/202039063}

\bibitem[{{Hikage} {et~al.}(2019){Hikage}, {Oguri}, {Hamana}, {More},
  {Mandelbaum}, {Takada}, {K{\"o}hlinger}, {Miyatake}, {Nishizawa}, {Aihara},
  {Armstrong}, {Bosch}, {Coupon}, {Ducout}, {Ho}, {Hsieh}, {Komiyama},
  {Lanusse}, {Leauthaud}, {Lupton}, {Medezinski}, {Mineo}, {Miyama},
  {Miyazaki}, {Murata}, {Murayama}, {Shirasaki}, {Sif{\'o}n}, {Simet},
  {Speagle}, {Spergel}, {Strauss}, {Sugiyama}, {Tanaka}, {Utsumi}, {Wang}, \&
  {Yamada}}]{2019PASJ...71...43H}
{Hikage}, C., {Oguri}, M., {Hamana}, T., {et~al.} 2019, \pasj, 71, 43,
  \dodoi{10.1093/pasj/psz010}

\bibitem[{{Hilbert} {et~al.}(2009{\natexlab{a}}){Hilbert}, {Hartlap}, {White},
  \& {Schneider}}]{2009A&A...499...31H}
{Hilbert}, S., {Hartlap}, J., {White}, S.~D.~M., \& {Schneider}, P.
  2009{\natexlab{a}}, \aap, 499, 31, \dodoi{10.1051/0004-6361/200811054}

\bibitem[{{Hilbert} {et~al.}(2009{\natexlab{b}}){Hilbert}, {Hartlap}, {White},
  \& {Schneider}}]{multi-lens}
---. 2009{\natexlab{b}}, \aap, 499, 31, \dodoi{10.1051/0004-6361/200811054}

\bibitem[{{Hildebrandt} {et~al.}(2017){Hildebrandt}, {Viola}, {Heymans},
  {Joudaki}, {Kuijken}, {Blake}, {Erben}, {Joachimi}, {Klaes}, {Miller},
  {Morrison}, {Nakajima}, {Verdoes Kleijn}, {Amon}, {Choi}, {Covone}, {de
  Jong}, {Dvornik}, {Fenech Conti}, {Grado}, {Harnois-D{\'e}raps}, {Herbonnet},
  {Hoekstra}, {K{\"o}hlinger}, {McFarland}, {Mead}, {Merten}, {Napolitano},
  {Peacock}, {Radovich}, {Schneider}, {Simon}, {Valentijn}, {van den Busch},
  {van Uitert}, \& {Van Waerbeke}}]{2017MNRAS.465.1454H}
{Hildebrandt}, H., {Viola}, M., {Heymans}, C., {et~al.} 2017, \mnras, 465,
  1454, \dodoi{10.1093/mnras/stw2805}

\bibitem[{{Hivon} {et~al.}(2002){Hivon}, {G{\'o}rski}, {Netterfield}, {Crill},
  {Prunet}, \& {Hansen}}]{pseudo2}
{Hivon}, E., {G{\'o}rski}, K.~M., {Netterfield}, C.~B., {et~al.} 2002, \apj,
  567, 2, \dodoi{10.1086/338126}

\bibitem[{{Hu} \& {Jain}(2004)}]{hu}
{Hu}, W., \& {Jain}, B. 2004, \prd, 70, 043009,
  \dodoi{10.1103/PhysRevD.70.043009}

\bibitem[{{Huterer}(2002)}]{2002PhRvD..65f3001H}
{Huterer}, D. 2002, \prd, 65, 063001, \dodoi{10.1103/PhysRevD.65.063001}

\bibitem[{{Huterer} {et~al.}(2006){Huterer}, {Takada}, {Bernstein}, \&
  {Jain}}]{2006MNRAS.366..101H}
{Huterer}, D., {Takada}, M., {Bernstein}, G., \& {Jain}, B. 2006, \mnras, 366,
  101, \dodoi{10.1111/j.1365-2966.2005.09782.x}

\bibitem[{{Ilbert} {et~al.}(2009){Ilbert}, {Capak}, {Salvato}, {Aussel},
  {McCracken}, {Sanders}, {Scoville}, {Kartaltepe}, {Arnouts}, {Le Floc'h},
  {Mobasher}, {Taniguchi}, {Lamareille}, {Leauthaud}, {Sasaki}, {Thompson},
  {Zamojski}, {Zamorani}, {Bardelli}, {Bolzonella}, {Bongiorno}, {Brusa},
  {Caputi}, {Carollo}, {Contini}, {Cook}, {Coppa}, {Cucciati}, {de la Torre},
  {de Ravel}, {Franzetti}, {Garilli}, {Hasinger}, {Iovino}, {Kampczyk},
  {Kneib}, {Knobel}, {Kovac}, {Le Borgne}, {Le Brun}, {Le F{\`e}vre}, {Lilly},
  {Looper}, {Maier}, {Mainieri}, {Mellier}, {Mignoli}, {Murayama}, {Pell{\`o}},
  {Peng}, {P{\'e}rez-Montero}, {Renzini}, {Ricciardelli}, {Schiminovich},
  {Scodeggio}, {Shioya}, {Silverman}, {Surace}, {Tanaka}, {Tasca}, {Tresse},
  {Vergani}, \& {Zucca}}]{2009ApJ...690.1236I}
{Ilbert}, O., {Capak}, P., {Salvato}, M., {et~al.} 2009, \apj, 690, 1236,
  \dodoi{10.1088/0004-637X/690/2/1236}

\bibitem[{{Ivezi{\'c}} {et~al.}(2019){Ivezi{\'c}}, {Kahn}, {Tyson}, {Abel},
  {Acosta}, {Allsman}, {Alonso}, {AlSayyad}, {Anderson}, {Andrew}, {Angel},
  {Angeli}, {Ansari}, {Antilogus}, {Araujo}, {Armstrong}, {Arndt}, {Astier},
  {Aubourg}, {Auza}, {Axelrod}, {Bard}, {Barr}, {Barrau}, {Bartlett}, {Bauer},
  {Bauman}, {Baumont}, {Bechtol}, {Bechtol}, {Becker}, {Becla}, {Beldica},
  {Bellavia}, {Bianco}, {Biswas}, {Blanc}, {Blazek}, {Blandford}, {Bloom},
  {Bogart}, {Bond}, {Booth}, {Borgland}, {Borne}, {Bosch}, {Boutigny},
  {Brackett}, {Bradshaw}, {Brandt}, {Brown}, {Bullock}, {Burchat}, {Burke},
  {Cagnoli}, {Calabrese}, {Callahan}, {Callen}, {Carlin}, {Carlson},
  {Chandrasekharan}, {Charles-Emerson}, {Chesley}, {Cheu}, {Chiang}, {Chiang},
  {Chirino}, {Chow}, {Ciardi}, {Claver}, {Cohen-Tanugi}, {Cockrum}, {Coles},
  {Connolly}, {Cook}, {Cooray}, {Covey}, {Cribbs}, {Cui}, {Cutri}, {Daly},
  {Daniel}, {Daruich}, {Daubard}, {Daues}, {Dawson}, {Delgado}, {Dellapenna},
  {de Peyster}, {de Val-Borro}, {Digel}, {Doherty}, {Dubois},
  {Dubois-Felsmann}, {Durech}, {Economou}, {Eifler}, {Eracleous}, {Emmons},
  {Fausti Neto}, {Ferguson}, {Figueroa}, {Fisher-Levine}, {Focke}, {Foss},
  {Frank}, {Freemon}, {Gangler}, {Gawiser}, {Geary}, {Gee}, {Geha}, {Gessner},
  {Gibson}, {Gilmore}, {Glanzman}, {Glick}, {Goldina}, {Goldstein}, {Goodenow},
  {Graham}, {Gressler}, {Gris}, {Guy}, {Guyonnet}, {Haller}, {Harris},
  {Hascall}, {Haupt}, {Hernandez}, {Herrmann}, {Hileman}, {Hoblitt}, {Hodgson},
  {Hogan}, {Howard}, {Huang}, {Huffer}, {Ingraham}, {Innes}, {Jacoby}, {Jain},
  {Jammes}, {Jee}, {Jenness}, {Jernigan}, {Jevremovi{\'c}}, {Johns}, {Johnson},
  {Johnson}, {Jones}, {Juramy-Gilles}, {Juri{\'c}}, {Kalirai}, {Kallivayalil},
  {Kalmbach}, {Kantor}, {Karst}, {Kasliwal}, {Kelly}, {Kessler}, {Kinnison},
  {Kirkby}, {Knox}, {Kotov}, {Krabbendam}, {Krughoff}, {Kub{\'a}nek},
  {Kuczewski}, {Kulkarni}, {Ku}, {Kurita}, {Lage}, {Lambert}, {Lange},
  {Langton}, {Le Guillou}, {Levine}, {Liang}, {Lim}, {Lintott}, {Long},
  {Lopez}, {Lotz}, {Lupton}, {Lust}, {MacArthur}, {Mahabal}, {Mandelbaum},
  {Markiewicz}, {Marsh}, {Marshall}, {Marshall}, {May}, {McKercher}, {McQueen},
  {Meyers}, {Migliore}, {Miller}, {Mills}, {Miraval}, {Moeyens}, {Moolekamp},
  {Monet}, {Moniez}, {Monkewitz}, {Montgomery}, {Morrison}, {Mueller},
  {Muller}, {Mu{\~n}oz Arancibia}, {Neill}, {Newbry}, {Nief}, {Nomerotski},
  {Nordby}, {O'Connor}, {Oliver}, {Olivier}, {Olsen}, {O'Mullane}, {Ortiz},
  {Osier}, {Owen}, {Pain}, {Palecek}, {Parejko}, {Parsons}, {Pease},
  {Peterson}, {Peterson}, {Petravick}, {Libby Petrick}, {Petry},
  {Pierfederici}, {Pietrowicz}, {Pike}, {Pinto}, {Plante}, {Plate}, {Plutchak},
  {Price}, {Prouza}, {Radeka}, {Rajagopal}, {Rasmussen}, {Regnault}, {Reil},
  {Reiss}, {Reuter}, {Ridgway}, {Riot}, {Ritz}, {Robinson}, {Roby}, {Roodman},
  {Rosing}, {Roucelle}, {Rumore}, {Russo}, {Saha}, {Sassolas}, {Schalk},
  {Schellart}, {Schindler}, {Schmidt}, {Schneider}, {Schneider}, {Schoening},
  {Schumacher}, {Schwamb}, {Sebag}, {Selvy}, {Sembroski}, {Seppala}, {Serio},
  {Serrano}, {Shaw}, {Shipsey}, {Sick}, {Silvestri}, {Slater}, {Smith},
  {Smith}, {Sobhani}, {Soldahl}, {Storrie-Lombardi}, {Stover}, {Strauss},
  {Street}, {Stubbs}, {Sullivan}, {Sweeney}, {Swinbank}, {Szalay}, {Takacs},
  {Tether}, {Thaler}, {Thayer}, {Thomas}, {Thornton}, {Thukral}, {Tice},
  {Trilling}, {Turri}, {Van Berg}, {Vanden Berk}, {Vetter}, {Virieux},
  {Vucina}, {Wahl}, {Walkowicz}, {Walsh}, {Walter}, {Wang}, {Wang}, {Warner},
  {Wiecha}, {Willman}, {Winters}, {Wittman}, {Wolff}, {Wood-Vasey}, {Wu},
  {Xin}, {Yoachim}, \& {Zhan}}]{LSST}
{Ivezi{\'c}}, {\v{Z}}., {Kahn}, S.~M., {Tyson}, J.~A., {et~al.} 2019, \apj,
  873, 111, \dodoi{10.3847/1538-4357/ab042c}

\bibitem[{{Joachimi} \& {Bridle}(2010)}]{3x2pt2}
{Joachimi}, B., \& {Bridle}, S.~L. 2010, \aap, 523, A1,
  \dodoi{10.1051/0004-6361/200913657}

\bibitem[{{Joachimi} {et~al.}(2015){Joachimi}, {Cacciato}, {Kitching},
  {Leonard}, {Mandelbaum}, {Sch{\"a}fer}, {Sif{\'o}n}, {Hoekstra}, {Kiessling},
  {Kirk}, \& {Rassat}}]{2015SSRv..193....1J}
{Joachimi}, B., {Cacciato}, M., {Kitching}, T.~D., {et~al.} 2015, \ssr, 193, 1,
  \dodoi{10.1007/s11214-015-0177-4}

\bibitem[{{Joachimi} {et~al.}(2021){Joachimi}, {Lin}, {Asgari}, {Tr{\"o}ster},
  {Heymans}, {Hildebrandt}, {K{\"o}hlinger}, {S{\'a}nchez}, {Wright},
  {Bilicki}, {Blake}, {van den Busch}, {Crocce}, {Dvornik}, {Erben}, {Getman},
  {Giblin}, {Hoekstra}, {Kannawadi}, {Kuijken}, {Napolitano}, {Schneider},
  {Scoccimarro}, {Sellentin}, {Shan}, {von Wietersheim-Kramsta}, \&
  {Zuntz}}]{KiDsfour3}
{Joachimi}, B., {Lin}, C.~A., {Asgari}, M., {et~al.} 2021, \aap, 646, A129,
  \dodoi{10.1051/0004-6361/202038831}

\bibitem[{{Joudaki} {et~al.}(2018){Joudaki}, {Blake}, {Johnson}, {Amon},
  {Asgari}, {Choi}, {Erben}, {Glazebrook}, {Harnois-D{\'e}raps}, {Heymans},
  {Hildebrandt}, {Hoekstra}, {Klaes}, {Kuijken}, {Lidman}, {Mead}, {Miller},
  {Parkinson}, {Poole}, {Schneider}, {Viola}, \& {Wolf}}]{2018MNRAS.474.4894J}
{Joudaki}, S., {Blake}, C., {Johnson}, A., {et~al.} 2018, \mnras, 474, 4894,
  \dodoi{10.1093/mnras/stx2820}

\bibitem[{{Kaiser}(1992)}]{1992ApJ...388..272K}
{Kaiser}, N. 1992, \apj, 388, 272, \dodoi{10.1086/171151}

\bibitem[{{Kaiser} \& {Squires}(1993)}]{KS}
{Kaiser}, N., \& {Squires}, G. 1993, \apj, 404, 441, \dodoi{10.1086/172297}

\bibitem[{{Kitching} {et~al.}(2012){Kitching}, {Balan}, {Bridle}, {Cantale},
  {Courbin}, {Eifler}, {Gentile}, {Gill}, {Harmeling}, {Heymans}, {Hirsch},
  {Honscheid}, {Kacprzak}, {Kirkby}, {Margala}, {Massey}, {Melchior},
  {Nurbaeva}, {Patton}, {Rhodes}, {Rowe}, {Taylor}, {Tewes}, {Viola},
  {Witherick}, {Voigt}, {Young}, \& {Zuntz}}]{2012MNRAS.423.3163K}
{Kitching}, T.~D., {Balan}, S.~T., {Bridle}, S., {et~al.} 2012, \mnras, 423,
  3163, \dodoi{10.1111/j.1365-2966.2012.21095.x}

\bibitem[{{Kitzbichler} \& {White}(2007)}]{2007MNRAS.376....2K}
{Kitzbichler}, M.~G., \& {White}, S.~D.~M. 2007, \mnras, 376, 2,
  \dodoi{10.1111/j.1365-2966.2007.11458.x}

\bibitem[{{Klypin} \& {Prada}(2019)}]{2019MNRAS.489.1684K}
{Klypin}, A., \& {Prada}, F. 2019, \mnras, 489, 1684,
  \dodoi{10.1093/mnras/stz2194}

\bibitem[{{Krause} {et~al.}(2017{\natexlab{a}}){Krause}, {Eifler}, {Zuntz},
  {Friedrich}, {Troxel}, {Dodelson}, {Blazek}, {Secco}, {MacCrann}, {Baxter},
  {Chang}, {Chen}, {Crocce}, {DeRose}, {Ferte}, {Kokron}, {Lacasa}, {Miranda},
  {Omori}, {Porredon}, {Rosenfeld}, {Samuroff}, {Wang}, {Wechsler}, {Abbott},
  {Abdalla}, {Allam}, {Annis}, {Bechtol}, {Benoit-Levy}, {Bernstein}, {Brooks},
  {Burke}, {Capozzi}, {Carrasco Kind}, {Carretero}, {D'Andrea}, {da Costa},
  {Davis}, {DePoy}, {Desai}, {Diehl}, {Dietrich}, {Evrard}, {Flaugher},
  {Fosalba}, {Frieman}, {Garcia-Bellido}, {Gaztanaga}, {Giannantonio}, {Gruen},
  {Gruendl}, {Gschwend}, {Gutierrez}, {Honscheid}, {James}, {Jeltema}, {Kuehn},
  {Kuhlmann}, {Lahav}, {Lima}, {Maia}, {March}, {Marshall}, {Martini},
  {Menanteau}, {Miquel}, {Nichol}, {Plazas}, {Romer}, {Rykoff}, {Sanchez},
  {Scarpine}, {Schindler}, {Schubnell}, {Sevilla-Noarbe}, {Smith},
  {Soares-Santos}, {Sobreira}, {Suchyta}, {Swanson}, {Tarle}, {Tucker},
  {Vikram}, {Walker}, \& {Weller}}]{3x2pt3}
{Krause}, E., {Eifler}, T.~F., {Zuntz}, J., {et~al.} 2017{\natexlab{a}}, arXiv
  e-prints, arXiv:1706.09359, \dodoi{10.48550/arXiv.1706.09359}

\bibitem[{{Krause} {et~al.}(2017{\natexlab{b}}){Krause}, {Eifler}, {Zuntz},
  {Friedrich}, {Troxel}, {Dodelson}, {Blazek}, {Secco}, {MacCrann}, {Baxter},
  {Chang}, {Chen}, {Crocce}, {DeRose}, {Ferte}, {Kokron}, {Lacasa}, {Miranda},
  {Omori}, {Porredon}, {Rosenfeld}, {Samuroff}, {Wang}, {Wechsler}, {Abbott},
  {Abdalla}, {Allam}, {Annis}, {Bechtol}, {Benoit-Levy}, {Bernstein}, {Brooks},
  {Burke}, {Capozzi}, {Carrasco Kind}, {Carretero}, {D'Andrea}, {da Costa},
  {Davis}, {DePoy}, {Desai}, {Diehl}, {Dietrich}, {Evrard}, {Flaugher},
  {Fosalba}, {Frieman}, {Garcia-Bellido}, {Gaztanaga}, {Giannantonio}, {Gruen},
  {Gruendl}, {Gschwend}, {Gutierrez}, {Honscheid}, {James}, {Jeltema}, {Kuehn},
  {Kuhlmann}, {Lahav}, {Lima}, {Maia}, {March}, {Marshall}, {Martini},
  {Menanteau}, {Miquel}, {Nichol}, {Plazas}, {Romer}, {Rykoff}, {Sanchez},
  {Scarpine}, {Schindler}, {Schubnell}, {Sevilla-Noarbe}, {Smith},
  {Soares-Santos}, {Sobreira}, {Suchyta}, {Swanson}, {Tarle}, {Tucker},
  {Vikram}, {Walker}, \& {Weller}}]{cov}
---. 2017{\natexlab{b}}, arXiv e-prints, arXiv:1706.09359,
  \dodoi{10.48550/arXiv.1706.09359}

\bibitem[{{Krause} {et~al.}(2021){Krause}, {Fang}, {Pandey}, {Secco}, {Alves},
  {Huang}, {Blazek}, {Prat}, {Zuntz}, {Eifler}, {MacCrann}, {DeRose}, {Crocce},
  {Porredon}, {Jain}, {Troxel}, {Dodelson}, {Huterer}, {Liddle}, {Leonard},
  {Amon}, {Chen}, {Elvin-Poole}, {Fert{\'e}}, {Muir}, {Park}, {Samuroff},
  {Brandao-Souza}, {Weaverdyck}, {Zacharegkas}, {Rosenfeld}, {Campos},
  {Chintalapati}, {Choi}, {Di Valentino}, {Doux}, {Herner}, {Lemos},
  {Mena-Fern{\'a}ndez}, {Omori}, {Paterno}, {Rodriguez-Monroy}, {Rogozenski},
  {Rollins}, {Troja}, {Tutusaus}, {Wechsler}, {Abbott}, {Aguena}, {Allam},
  {Andrade-Oliveira}, {Annis}, {Bacon}, {Baxter}, {Bechtol}, {Bernstein},
  {Brooks}, {Buckley-Geer}, {Burke}, {Carnero Rosell}, {Carrasco Kind},
  {Carretero}, {Castander}, {Cawthon}, {Chang}, {Costanzi}, {da Costa},
  {Pereira}, {De Vicente}, {Desai}, {Diehl}, {Doel}, {Everett}, {Evrard},
  {Ferrero}, {Flaugher}, {Fosalba}, {Frieman}, {Garc{\'\i}a-Bellido},
  {Gaztanaga}, {Gerdes}, {Giannantonio}, {Gruen}, {Gruendl}, {Gschwend},
  {Gutierrez}, {Hartley}, {Hinton}, {Hollowood}, {Honscheid}, {Hoyle}, {Huff},
  {James}, {Kuehn}, {Kuropatkin}, {Lahav}, {Lima}, {Maia}, {Marshall},
  {Martini}, {Melchior}, {Menanteau}, {Miquel}, {Mohr}, {Morgan}, {Myles},
  {Palmese}, {Paz-Chinch{\'o}n}, {Petravick}, {Pieres}, {Plazas Malag{\'o}n},
  {Sanchez}, {Scarpine}, {Schubnell}, {Serrano}, {Sevilla-Noarbe}, {Smith},
  {Soares-Santos}, {Suchyta}, {Tarle}, {Thomas}, {To}, {Varga}, \&
  {Weller}}]{3x2pt4}
{Krause}, E., {Fang}, X., {Pandey}, S., {et~al.} 2021, arXiv e-prints,
  arXiv:2105.13548, \dodoi{10.48550/arXiv.2105.13548}

\bibitem[{{Lewis} {et~al.}(2000){Lewis}, {Challinor}, \& {Lasenby}}]{camb}
{Lewis}, A., {Challinor}, A., \& {Lasenby}, A. 2000, \apj, 538, 473,
  \dodoi{10.1086/309179}

\bibitem[{{Limber}(1954)}]{Limber}
{Limber}, D.~N. 1954, \apj, 119, 655, \dodoi{10.1086/145870}

\bibitem[{{Lin} {et~al.}(2024){Lin}, {Deng}, {Gong}, \& {Chen}}]{LinAxion}
{Lin}, H., {Deng}, F., {Gong}, Y., \& {Chen}, X. 2024, \mnras, 529, 1542,
  \dodoi{10.1093/mnras/stae627}

\bibitem[{{Lin} {et~al.}(2022){Lin}, {Gong}, {Chen}, {Chan}, {Fan}, \&
  {Zhan}}]{lin}
{Lin}, H., {Gong}, Y., {Chen}, X., {et~al.} 2022, \mnras, 515, 5743,
  \dodoi{10.1093/mnras/stac2126}

\bibitem[{{Looijmans} {et~al.}(2024){Looijmans}, {Wang}, \&
  {Beutler}}]{2024arXiv240213783L}
{Looijmans}, M.~J., {Wang}, M.~S., \& {Beutler}, F. 2024, arXiv e-prints,
  arXiv:2402.13783, \dodoi{10.48550/arXiv.2402.13783}

\bibitem[{Massey {et~al.}(2013)Massey, Hoekstra, Kitching, Rhodes, Cropper,
  Amiaux, Harvey, Mellier, Meneghetti, Miller, {et~al.}}]{massey2013origins}
Massey, R., Hoekstra, H., Kitching, T., {et~al.} 2013, Monthly Notices of the
  Royal Astronomical Society, 429, 661

\bibitem[{{Miao} {et~al.}(2023){Miao}, {Gong}, {Chen}, {Huang}, {Li}, \&
  {Zhan}}]{Miao}
{Miao}, H., {Gong}, Y., {Chen}, X., {et~al.} 2023, \mnras, 519, 1132,
  \dodoi{10.1093/mnras/stac3583}

\bibitem[{{Murray}(2018)}]{pbox}
{Murray}, S.~G. 2018, {powerbox: Arbitrarily structured, arbitrary-dimension
  boxes and log-normal mocks}, Astrophysics Source Code Library, record
  ascl:1805.001

\bibitem[{{Navarro} {et~al.}(1996){Navarro}, {Frenk}, \& {White}}]{NFW}
{Navarro}, J.~F., {Frenk}, C.~S., \& {White}, S. D.~M. 1996, \apj, 462, 563,
  \dodoi{10.1086/177173}

\bibitem[{{Pandey} {et~al.}(2022){Pandey}, {Krause}, {DeRose}, {MacCrann},
  {Jain}, {Crocce}, {Blazek}, {Choi}, {Huang}, {To}, {Fang}, {Elvin-Poole},
  {Prat}, {Porredon}, {Secco}, {Rodriguez-Monroy}, {Weaverdyck}, {Park},
  {Raveri}, {Rozo}, {Rykoff}, {Bernstein}, {S{\'a}nchez}, {Jarvis}, {Troxel},
  {Zacharegkas}, {Chang}, {Alarcon}, {Alves}, {Amon}, {Andrade-Oliveira},
  {Baxter}, {Bechtol}, {Becker}, {Camacho}, {Campos}, {Carnero Rosell},
  {Carrasco Kind}, {Cawthon}, {Chen}, {Chintalapati}, {Davis}, {Di Valentino},
  {Diehl}, {Dodelson}, {Doux}, {Drlica-Wagner}, {Eckert}, {Eifler}, {Elsner},
  {Everett}, {Farahi}, {Fert{\'e}}, {Fosalba}, {Friedrich}, {Gatti},
  {Giannini}, {Gruen}, {Gruendl}, {Harrison}, {Hartley}, {Huff}, {Huterer},
  {Kovacs}, {Leget}, {McCullough}, {Muir}, {Myles}, {Navarro-Alsina}, {Omori},
  {Rollins}, {Roodman}, {Rosenfeld}, {Sevilla-Noarbe}, {Sheldon}, {Shin},
  {Troja}, {Tutusaus}, {Varga}, {Wechsler}, {Yanny}, {Yin}, {Zhang}, {Zuntz},
  {Abbott}, {Aguena}, {Allam}, {Annis}, {Bacon}, {Bertin}, {Brooks}, {Burke},
  {Carretero}, {Conselice}, {Costanzi}, {da Costa}, {Pereira}, {De Vicente},
  {Dietrich}, {Doel}, {Evrard}, {Ferrero}, {Flaugher}, {Frieman},
  {Garc{\'\i}a-Bellido}, {Gaztanaga}, {Gerdes}, {Giannantonio}, {Gschwend},
  {Gutierrez}, {Hinton}, {Hollowood}, {Honscheid}, {James}, {Jeltema}, {Kuehn},
  {Kuropatkin}, {Lahav}, {Lima}, {Lin}, {Maia}, {Marshall}, {Melchior},
  {Menanteau}, {Miller}, {Miquel}, {Mohr}, {Morgan}, {Palmese},
  {Paz-Chinch{\'o}n}, {Petravick}, {Pieres}, {Plazas Malag{\'o}n}, {Sanchez},
  {Scarpine}, {Serrano}, {Smith}, {Soares-Santos}, {Suchyta}, {Tarle},
  {Thomas}, {Weller}, \& {DES Collaboration}}]{2022PhRvD.106d3520P}
{Pandey}, S., {Krause}, E., {DeRose}, J., {et~al.} 2022, \prd, 106, 043520,
  \dodoi{10.1103/PhysRevD.106.043520}

\bibitem[{{Peebles}(1973)}]{1973ApJ...185..413P}
{Peebles}, P.~J.~E. 1973, \apj, 185, 413, \dodoi{10.1086/152431}

\bibitem[{{Peebles}(2012)}]{LCDM2}
---. 2012, \araa, 50, 1, \dodoi{10.1146/annurev-astro-081811-125526}

\bibitem[{{Pei} {et~al.}(2024){Pei}, {Guo}, {Li}, {Wang}, {Han}, {Hu}, {Su},
  {Gao}, {Wang}, {Luo}, \& {Wei}}]{2024MNRAS.529.4958P}
{Pei}, W., {Guo}, Q., {Li}, M., {et~al.} 2024, \mnras, 529, 4958,
  \dodoi{10.1093/mnras/stae866}

\bibitem[{{Petri}(2016)}]{lenstools}
{Petri}, A. 2016, Astronomy and Computing, 17, 73,
  \dodoi{10.1016/j.ascom.2016.06.001}

\bibitem[{{Pires} {et~al.}(2010){Pires}, {Starck}, {Amara}, {Teyssier},
  {Refregier}, \& {Fadili}}]{inpainting2}
{Pires}, S., {Starck}, J.~L., {Amara}, A., {et~al.} 2010, {FASTLens (FAst
  STatistics for weak Lensing): Fast Method for Weak Lensing Statistics and Map
  Making}, Astrophysics Source Code Library, record ascl:1010.041

\bibitem[{{Planck Collaboration} {et~al.}(2020){Planck Collaboration},
  {Aghanim}, {Akrami}, {Ashdown}, {Aumont}, {Baccigalupi}, {Ballardini},
  {Banday}, {Barreiro}, {Bartolo}, {Basak}, {Battye}, {Benabed}, {Bernard},
  {Bersanelli}, {Bielewicz}, {Bock}, {Bond}, {Borrill}, {Bouchet}, {Boulanger},
  {Bucher}, {Burigana}, {Butler}, {Calabrese}, {Cardoso}, {Carron},
  {Challinor}, {Chiang}, {Chluba}, {Colombo}, {Combet}, {Contreras}, {Crill},
  {Cuttaia}, {de Bernardis}, {de Zotti}, {Delabrouille}, {Delouis}, {Di
  Valentino}, {Diego}, {Dor{\'e}}, {Douspis}, {Ducout}, {Dupac}, {Dusini},
  {Efstathiou}, {Elsner}, {En{\ss}lin}, {Eriksen}, {Fantaye}, {Farhang},
  {Fergusson}, {Fernandez-Cobos}, {Finelli}, {Forastieri}, {Frailis},
  {Fraisse}, {Franceschi}, {Frolov}, {Galeotta}, {Galli}, {Ganga},
  {G{\'e}nova-Santos}, {Gerbino}, {Ghosh}, {Gonz{\'a}lez-Nuevo}, {G{\'o}rski},
  {Gratton}, {Gruppuso}, {Gudmundsson}, {Hamann}, {Handley}, {Hansen},
  {Herranz}, {Hildebrandt}, {Hivon}, {Huang}, {Jaffe}, {Jones}, {Karakci},
  {Keih{\"a}nen}, {Keskitalo}, {Kiiveri}, {Kim}, {Kisner}, {Knox},
  {Krachmalnicoff}, {Kunz}, {Kurki-Suonio}, {Lagache}, {Lamarre}, {Lasenby},
  {Lattanzi}, {Lawrence}, {Le Jeune}, {Lemos}, {Lesgourgues}, {Levrier},
  {Lewis}, {Liguori}, {Lilje}, {Lilley}, {Lindholm}, {L{\'o}pez-Caniego},
  {Lubin}, {Ma}, {Mac{\'\i}as-P{\'e}rez}, {Maggio}, {Maino}, {Mandolesi},
  {Mangilli}, {Marcos-Caballero}, {Maris}, {Martin}, {Martinelli},
  {Mart{\'\i}nez-Gonz{\'a}lez}, {Matarrese}, {Mauri}, {McEwen}, {Meinhold},
  {Melchiorri}, {Mennella}, {Migliaccio}, {Millea}, {Mitra},
  {Miville-Desch{\^e}nes}, {Molinari}, {Montier}, {Morgante}, {Moss}, {Natoli},
  {N{\o}rgaard-Nielsen}, {Pagano}, {Paoletti}, {Partridge}, {Patanchon},
  {Peiris}, {Perrotta}, {Pettorino}, {Piacentini}, {Polastri}, {Polenta},
  {Puget}, {Rachen}, {Reinecke}, {Remazeilles}, {Renzi}, {Rocha}, {Rosset},
  {Roudier}, {Rubi{\~n}o-Mart{\'\i}n}, {Ruiz-Granados}, {Salvati}, {Sandri},
  {Savelainen}, {Scott}, {Shellard}, {Sirignano}, {Sirri}, {Spencer},
  {Sunyaev}, {Suur-Uski}, {Tauber}, {Tavagnacco}, {Tenti}, {Toffolatti},
  {Tomasi}, {Trombetti}, {Valenziano}, {Valiviita}, {Van Tent}, {Vibert},
  {Vielva}, {Villa}, {Vittorio}, {Wandelt}, {Wehus}, {White}, {White},
  {Zacchei}, \& {Zonca}}]{planck18}
{Planck Collaboration}, {Aghanim}, N., {Akrami}, Y., {et~al.} 2020, \aap, 641,
  A6, \dodoi{10.1051/0004-6361/201833910}

\bibitem[{{Porredon} {et~al.}(2022{\natexlab{a}}){Porredon}, {Crocce},
  {Elvin-Poole}, {Cawthon}, {Giannini}, {De Vicente}, {Carnero Rosell},
  {Ferrero}, {Krause}, {Fang}, {Prat}, {Rodriguez-Monroy}, {Pandey}, {Pocino},
  {Castander}, {Choi}, {Amon}, {Tutusaus}, {Dodelson}, {Sevilla-Noarbe},
  {Fosalba}, {Gaztanaga}, {Alarcon}, {Alves}, {Andrade-Oliveira}, {Baxter},
  {Bechtol}, {Becker}, {Bernstein}, {Blazek}, {Camacho}, {Campos}, {Carrasco
  Kind}, {Chintalapati}, {Cordero}, {DeRose}, {Di Valentino}, {Doux}, {Eifler},
  {Everett}, {Fert{\'e}}, {Friedrich}, {Gatti}, {Gruen}, {Harrison}, {Hartley},
  {Herner}, {Huff}, {Huterer}, {Jain}, {Jarvis}, {Lee}, {Lemos}, {MacCrann},
  {Mena-Fern{\'a}ndez}, {Muir}, {Myles}, {Park}, {Raveri}, {Rosenfeld}, {Ross},
  {Rykoff}, {Samuroff}, {S{\'a}nchez}, {Sanchez}, {Sanchez}, {Sanchez Cid},
  {Scolnic}, {Secco}, {Sheldon}, {Troja}, {Troxel}, {Weaverdyck}, {Yanny},
  {Zuntz}, {Abbott}, {Aguena}, {Allam}, {Annis}, {Avila}, {Bacon}, {Bertin},
  {Bhargava}, {Brooks}, {Buckley-Geer}, {Burke}, {Carretero}, {Costanzi}, {da
  Costa}, {Pereira}, {Davis}, {Desai}, {Diehl}, {Dietrich}, {Doel},
  {Drlica-Wagner}, {Eckert}, {Evrard}, {Flaugher}, {Frieman},
  {Garc{\'\i}a-Bellido}, {Gerdes}, {Giannantonio}, {Gruendl}, {Gschwend},
  {Gutierrez}, {Hinton}, {Hollowood}, {Honscheid}, {Hoyle}, {James}, {Kuehn},
  {Kuropatkin}, {Lahav}, {Lidman}, {Lima}, {Lin}, {Maia}, {Marshall},
  {Martini}, {Melchior}, {Menanteau}, {Miquel}, {Mohr}, {Morgan}, {Ogando},
  {Palmese}, {Paz-Chinch{\'o}n}, {Petravick}, {Pieres}, {Plazas Malag{\'o}n},
  {Romer}, {Santiago}, {Scarpine}, {Schubnell}, {Serrano}, {Smith},
  {Soares-Santos}, {Suchyta}, {Tarle}, {Thomas}, {To}, {Varga}, {Weller}, \&
  {DES Collaboration}}]{redun2}
{Porredon}, A., {Crocce}, M., {Elvin-Poole}, J., {et~al.} 2022{\natexlab{a}},
  \prd, 106, 103530, \dodoi{10.1103/PhysRevD.106.103530}

\bibitem[{{Porredon} {et~al.}(2022{\natexlab{b}}){Porredon}, {Crocce},
  {Elvin-Poole}, {Cawthon}, {Giannini}, {De Vicente}, {Carnero Rosell},
  {Ferrero}, {Krause}, {Fang}, {Prat}, {Rodriguez-Monroy}, {Pandey}, {Pocino},
  {Castander}, {Choi}, {Amon}, {Tutusaus}, {Dodelson}, {Sevilla-Noarbe},
  {Fosalba}, {Gaztanaga}, {Alarcon}, {Alves}, {Andrade-Oliveira}, {Baxter},
  {Bechtol}, {Becker}, {Bernstein}, {Blazek}, {Camacho}, {Campos}, {Carrasco
  Kind}, {Chintalapati}, {Cordero}, {DeRose}, {Di Valentino}, {Doux}, {Eifler},
  {Everett}, {Fert{\'e}}, {Friedrich}, {Gatti}, {Gruen}, {Harrison}, {Hartley},
  {Herner}, {Huff}, {Huterer}, {Jain}, {Jarvis}, {Lee}, {Lemos}, {MacCrann},
  {Mena-Fern{\'a}ndez}, {Muir}, {Myles}, {Park}, {Raveri}, {Rosenfeld}, {Ross},
  {Rykoff}, {Samuroff}, {S{\'a}nchez}, {Sanchez}, {Sanchez}, {Sanchez Cid},
  {Scolnic}, {Secco}, {Sheldon}, {Troja}, {Troxel}, {Weaverdyck}, {Yanny},
  {Zuntz}, {Abbott}, {Aguena}, {Allam}, {Annis}, {Avila}, {Bacon}, {Bertin},
  {Bhargava}, {Brooks}, {Buckley-Geer}, {Burke}, {Carretero}, {Costanzi}, {da
  Costa}, {Pereira}, {Davis}, {Desai}, {Diehl}, {Dietrich}, {Doel},
  {Drlica-Wagner}, {Eckert}, {Evrard}, {Flaugher}, {Frieman},
  {Garc{\'\i}a-Bellido}, {Gerdes}, {Giannantonio}, {Gruendl}, {Gschwend},
  {Gutierrez}, {Hinton}, {Hollowood}, {Honscheid}, {Hoyle}, {James}, {Kuehn},
  {Kuropatkin}, {Lahav}, {Lidman}, {Lima}, {Lin}, {Maia}, {Marshall},
  {Martini}, {Melchior}, {Menanteau}, {Miquel}, {Mohr}, {Morgan}, {Ogando},
  {Palmese}, {Paz-Chinch{\'o}n}, {Petravick}, {Pieres}, {Plazas Malag{\'o}n},
  {Romer}, {Santiago}, {Scarpine}, {Schubnell}, {Serrano}, {Smith},
  {Soares-Santos}, {Suchyta}, {Tarle}, {Thomas}, {To}, {Varga}, {Weller}, \&
  {DES Collaboration}}]{comres2}
---. 2022{\natexlab{b}}, \prd, 106, 103530, \dodoi{10.1103/PhysRevD.106.103530}

\bibitem[{{Riess} {et~al.}(2022){Riess}, {Yuan}, {Macri}, {Scolnic}, {Brout},
  {Casertano}, {Jones}, {Murakami}, {Anand}, {Breuval}, {Brink}, {Filippenko},
  {Hoffmann}, {Jha}, {D'arcy Kenworthy}, {Mackenty}, {Stahl}, \&
  {Zheng}}]{typeIa1}
{Riess}, A.~G., {Yuan}, W., {Macri}, L.~M., {et~al.} 2022, \apjl, 934, L7,
  \dodoi{10.3847/2041-8213/ac5c5b}

\bibitem[{{Schaan} {et~al.}(2020){Schaan}, {Ferraro}, \&
  {Seljak}}]{2020JCAP...12..001S}
{Schaan}, E., {Ferraro}, S., \& {Seljak}, U. 2020, \jcap, 2020, 001,
  \dodoi{10.1088/1475-7516/2020/12/001}

\bibitem[{Schneider {et~al.}(2016)Schneider, Teyssier, Potter, Stadel, Onions,
  Reed, Smith, Springel, Pearce, \& Scoccimarro}]{schneider2016matter}
Schneider, A., Teyssier, R., Potter, D., {et~al.} 2016, Journal of Cosmology
  and Astroparticle Physics, 2016, 047

\bibitem[{{Shirasaki} {et~al.}(2013){Shirasaki}, {Yoshida}, \&
  {Hamana}}]{masked}
{Shirasaki}, M., {Yoshida}, N., \& {Hamana}, T. 2013, \apj, 774, 111,
  \dodoi{10.1088/0004-637X/774/2/111}

\bibitem[{{Smith} {et~al.}(2020){Smith}, {Sullivan}, {Wiseman}, {Kessler},
  {Scolnic}, {Brout}, {D'Andrea}, {Davis}, {Foley}, {Frohmaier}, {Galbany},
  {Gupta}, {Guti{\'e}rrez}, {Hinton}, {Kelsey}, {Lidman}, {Macaulay},
  {M{\"o}ller}, {Nichol}, {Nugent}, {Palmese}, {Pursiainen}, {Sako}, {Swann},
  {Thomas}, {Tucker}, {Vincenzi}, {Carollo}, {Lewis}, {Sommer}, {Abbott},
  {Aguena}, {Allam}, {Avila}, {Bertin}, {Bhargava}, {Brooks}, {Buckley-Geer},
  {Burke}, {Carnero Rosell}, {Carrasco Kind}, {Costanzi}, {da Costa}, {De
  Vicente}, {Desai}, {Diehl}, {Doel}, {Eifler}, {Everett}, {Flaugher},
  {Fosalba}, {Frieman}, {Garc{\'\i}a-Bellido}, {Gaztanaga}, {Glazebrook},
  {Gruen}, {Gruendl}, {Gschwend}, {Gutierrez}, {Hartley}, {Hollowood},
  {Honscheid}, {James}, {Krause}, {Kuehn}, {Kuropatkin}, {Lima}, {MacCrann},
  {Maia}, {Marshall}, {Martini}, {Melchior}, {Menanteau}, {Miquel},
  {Paz-Chinch{\'o}n}, {Plazas}, {Romer}, {Roodman}, {Rykoff}, {Sanchez},
  {Scarpine}, {Schubnell}, {Serrano}, {Sevilla-Noarbe}, {Suchyta}, {Swanson},
  {Tarle}, {Thomas}, {Tucker}, {Varga}, {Walker}, \& {DES
  Collaboration}}]{typeIa2}
{Smith}, M., {Sullivan}, M., {Wiseman}, P., {et~al.} 2020, \mnras, 494, 4426,
  \dodoi{10.1093/mnras/staa946}

\bibitem[{{Springel}(2005)}]{2005MNRAS.364.1105S}
{Springel}, V. 2005, \mnras, 364, 1105,
  \dodoi{10.1111/j.1365-2966.2005.09655.x}

\bibitem[{{Springel} {et~al.}(2001){Springel}, {Yoshida}, \&
  {White}}]{2001NewA....6...79S}
{Springel}, V., {Yoshida}, N., \& {White}, S. D.~M. 2001, \na, 6, 79,
  \dodoi{10.1016/S1384-1076(01)00042-2}

\bibitem[{{Takada} \& {Hu}(2013)}]{SSC}
{Takada}, M., \& {Hu}, W. 2013, \prd, 87, 123504,
  \dodoi{10.1103/PhysRevD.87.123504}

\bibitem[{{Takada} \& {Jain}(2004)}]{cNG}
{Takada}, M., \& {Jain}, B. 2004, \mnras, 348, 897,
  \dodoi{10.1111/j.1365-2966.2004.07410.x}

\bibitem[{{Takahashi} {et~al.}(2012){Takahashi}, {Sato}, {Nishimichi},
  {Taruya}, \& {Oguri}}]{Taka}
{Takahashi}, R., {Sato}, M., {Nishimichi}, T., {Taruya}, A., \& {Oguri}, M.
  2012, \apj, 761, 152, \dodoi{10.1088/0004-637X/761/2/152}

\bibitem[{{Tinker} {et~al.}(2008){Tinker}, {Kravtsov}, {Klypin}, {Abazajian},
  {Warren}, {Yepes}, {Gottl{\"o}ber}, \& {Holz}}]{tink08}
{Tinker}, J., {Kravtsov}, A.~V., {Klypin}, A., {et~al.} 2008, \apj, 688, 709,
  \dodoi{10.1086/591439}

\bibitem[{{Tinker} {et~al.}(2010){Tinker}, {Robertson}, {Kravtsov}, {Klypin},
  {Warren}, {Yepes}, \& {Gottl{\"o}ber}}]{tink10}
{Tinker}, J.~L., {Robertson}, B.~E., {Kravtsov}, A.~V., {et~al.} 2010, \apj,
  724, 878, \dodoi{10.1088/0004-637X/724/2/878}

\bibitem[{{van Uitert} {et~al.}(2018){van Uitert}, {Joachimi}, {Joudaki},
  {Amon}, {Heymans}, {K{\"o}hlinger}, {Asgari}, {Blake}, {Choi}, {Erben},
  {Farrow}, {Harnois-D{\'e}raps}, {Hildebrandt}, {Hoekstra}, {Kitching},
  {Klaes}, {Kuijken}, {Merten}, {Miller}, {Nakajima}, {Schneider}, {Valentijn},
  \& {Viola}}]{KiDsfour1}
{van Uitert}, E., {Joachimi}, B., {Joudaki}, S., {et~al.} 2018, \mnras, 476,
  4662, \dodoi{10.1093/mnras/sty551}

\bibitem[{{Van Waerbeke} {et~al.}(2000){Van Waerbeke}, {Mellier}, {Erben},
  {Cuillandre}, {Bernardeau}, {Maoli}, {Bertin}, {McCracken}, {Le F{\`e}vre},
  {Fort}, {Dantel-Fort}, {Jain}, \& {Schneider}}]{2000A&A...358...30V}
{Van Waerbeke}, L., {Mellier}, Y., {Erben}, T., {et~al.} 2000, \aap, 358, 30,
  \dodoi{10.48550/arXiv.astro-ph/0002500}

\bibitem[{{Yao} {et~al.}(2024){Yao}, {Shan}, {Li}, {Xu}, {Fan}, {Liu}, {Zhang},
  {Yu}, {Wei}, {Hu}, {Li}, {Fan}, {Xu}, \& {Guo}}]{yao}
{Yao}, J., {Shan}, H., {Li}, R., {et~al.} 2024, \mnras, 527, 5206,
  \dodoi{10.1093/mnras/stad3563}

\bibitem[{{Zhan}(2011)}]{zhan1}
{Zhan}, H. 2011, Scientia Sinica Physica, Mechanica \& Astronomica, 41, 1441,
  \dodoi{10.1360/132011-961}

\bibitem[{{Zhan}(2018)}]{zhan2}
{Zhan}, H. 2018, in 42nd COSPAR Scientific Assembly, Vol.~42, E1.16--4--18

\bibitem[{Zhan(2021)}]{zhan3}
Zhan, H. 2021, Chinese Science Bulletin, 66, 1290, \dodoi{10.1360/TB-2021-0016}

\bibitem[{{Zorrilla Matilla} {et~al.}(2020){Zorrilla Matilla}, {Waterval}, \&
  {Haiman}}]{2020AJ....159..284Z}
{Zorrilla Matilla}, J.~M., {Waterval}, S., \& {Haiman}, Z. 2020, \aj, 159, 284,
  \dodoi{10.3847/1538-3881/ab8f8c}

\end{thebibliography}
\bibliographystyle{aasjournal}



\end{document}